\documentclass[draft]{agujournal2019}
\usepackage{url}
\usepackage{lineno}
\usepackage[inline]{trackchanges}
\usepackage{soul}
\usepackage{setspace}
\usepackage{amsmath}
\usepackage{amssymb}
\usepackage{graphicx}

\draftfalse

\journalname{AGU Advances}

\begin{document}
\justifying
%
%


\title{Major Space Weather Risks Identified via Coupled Physics-Engineering-Economic Modeling}

%
%




\authors{
	Edward~J.~Oughton$^{1\ast\dagger}$,
	Dennies~K.~Bor$^{1\dagger}$,
	Robert~S.~Weigel$^{1}$, \\
	C.~Trevor~Gaunt$^{2}$, 
	Ridvan~Dogan$^{1}$,
	Liling~Huang$^{1}$, \\
	Jeffrey~J.~Love$^{3}$,
	Michael~Wiltberger$^{4}$  \\
	\small$^{1}$Space Weather Lab, George Mason University, Fairfax, VA, USA. \\
	\small$^{2}$University of Cape Town, Cape Town, South Africa.\\
	\small$^{3}$U.S. Geological Survey, Geologic Hazards Science Center, Denver, CO, USA.\\
	\small$^{4}$U.S. National Science Foundation National Center for Atmospheric Research, Boulder, CO, USA. \\
	\small$^\ast$Corresponding author (email: eoughton@gmu.edu).
	\small$^\dagger$Authors contributed equally.
}









\begin{keypoints}
\item Space weather poses an important but under-quantified threat to society.
\item A novel physics-engineering-economic framework assesses power grid socio-economic impacts with quantified uncertainties.
\item A daily U.S. loss of 2.04 billion USD is estimated for a 250-year geomagnetic storm (95 percent confidence interval: 1.86 to 2.22 billion USD).
\end{keypoints}

\vfill
\noindent\textit{\small\textbf{Peer Review DISCLAIMER:} This draft manuscript is distributed solely for purposes of scientific peer review. Its content is deliberative and predecisional, so it must not be disclosed or released by reviewers. Because the manuscript has not yet been approved for publication by the U.S. Geological Survey (USGS), it does not represent any official USGS finding or policy.}
\vspace{1em}

%
%

%
%

\newpage

\section*{Abstract} 
Space weather poses an important but under-quantified threat to society. While severe geomagnetic storms are recognized as potential global catastrophes, their socio-economic impacts remain poorly quantified. We present a novel physics-engineering-economic framework that links geophysical drivers to power grid geoelectric fields, transformer vulnerability, and macroeconomic consequences. Using the United States as an example, we estimate daily U.S. economic losses for a 250-year geomagnetic storm from transformer thermal heating of 2.04 billion USD (95 percent confidence interval: 1.86 to 2.22 billion USD), disrupting power for approximately 5.7 million people and 150,000 businesses. These estimates are conservative lower bounds, reflecting only transformer thermal heating effects and excluding voltage collapse, cascading failures, and restoration costs. The true societal risk is likely substantially higher. Nonetheless, the contribution is in providing the first nationwide end-to-end coupling from space physics to potential macroeconomic loss, with quantified uncertainties. Our results demonstrate that coupled socio-economic modeling of space weather is both feasible and essential, and the framework is scalable and transferable, offering a template for assessing space weather risk to critical infrastructure in other countries.
\newpage

\section{Introduction}
\noindent
Large coronal mass ejections (CMEs) and associated geomagnetic storms induce geoelectric fields that can drive quasi-direct currents through extra-high-voltage (EHV) transformers, affecting protective relays, and destabilizing power systems \cite{gopalswamy_sun_2022}. The resulting consequences of critical infrastructure failure affect households and businesses due to disruptions in essential digital services, health care, finance, manufacturing, logistics, and government operations \cite{bhattarai_ensemble_2025, oughton_stochastic_2019, montoya-rincon_socio-technical_2023}. 

Despite decades of advances in space physics and power engineering, the socio-economic consequences of severe space weather remain uncertain because the key disciplines are often studied in isolation \cite{oughton_risk_2019}. Consequently, policy makers and critical infrastructure operators have had to consider mitigation options without a common, evidence-based approach to translating storm scenarios into population impacts and economic losses. Generally, the options considered range from introducing basic geomagnetically induced current (GIC) monitoring and potentially new operational procedures, to more capital-intensive investments such as grid islanding, GIC blocking, or asset replacement \cite{rajput_insight_2021, mac_manus_geomagnetically_2023, kappenman_gic_1991}.

Existing economic assessments are frequently deterministic or scenario-based, often omitting the tight coupling between geophysical drivers, induced geoelectric fields, substation-specific exposure, and transformer failure mechanisms \cite{oughton_quantifying_2017, schulte_how_2014, ishii_space_2021, abt_associates_social_2017, eastwood_quantifying_2018}. Conversely, technically rigorous studies have tended to focus on either the physics of geoelectric hazards \cite{love_geoelectric_2018, love_empirical_2019, pulkkinen_statistics_2008, marshalko_three-dimensional_2023} or the engineering of grid responses \cite{hughes_revealing_2022, gaunt_why_2016, overbye_power_2013, birchfield_statistical_2017, juvekar_gic-inclusive_2021}, rarely extending to downstream socio-economic analysis, which is essential to support business and policy decisions. This issue urgently needs to be addressed, so stakeholders can access quantitative estimates of socio-economic risk with associated uncertainties, properly reflecting the underlying physics and engineering dimensions of this hazard. Thus, responding to priorities articulated in national space weather strategies, such as the U.S. National Space Weather Strategy and Action Plan \cite{sworm_national_2019, sworm_implementation_2023}, we bridge this gap with an end-to-end framework that couples geophysical drivers, grid exposure and reliability, and economic propagation into a single value-at-risk assessment.

Our contribution is threefold. First, we develop a nationally scalable modular framework that couples (i) spatial event data for the geoelectric hazard, (ii) a bulk power transmission network model, incorporating high-voltage and EHV assets ($\geq$161 kV), with probabilistic transformer configurations and age-dependent fragility, and (iii) a macroeconomic model that converts localized outages into daily potential sectoral and national losses. Second, we validate the framework against utility measurements in the Tennessee Valley Authority network during the recent ``Gannon'' storm of May 2024 \cite{wilkerson_gic--related_2025}, demonstrating fidelity at the time and frequency scales most responsible for transformer heating and GIC exposure. Third, focusing on the United States, we quantify risk across events of increasing severity, providing uncertainty estimates for impacts on the population and businesses potentially affected, along with direct and total economic losses. The framework's data and model interfaces make it transferable to other countries, enabling a global template for space weather risk assessment to support policy decisions.

\subsection{Thermal heating in transformers}\label{intro:thermal-heating}
        
Electricity transmission systems, specifically those involving EHV assets, are highly vulnerable to GICs \cite{dehghanian_integrated_2024,overbye_integration_2012,boteler_modeling_2017,khurshid_impact_2021}. During severe geomagnetic storms, GICs are driven by rapid temporal and spatial changes in Earth's magnetic field, which are in turn caused by interactions between Earth's magnetosphere and southward-oriented CMEs \cite{ganushkina_current_2018,hajra_intense_2022}. These interactions initiate magnetic reconnection processes that inject energetic charged particles into the magnetosphere, enhancing ring currents and causing ionospheric disturbances \cite{banerjee_existence_2012,love_mapping_2022,stevens_generating_2023}.
The resulting geomagnetic fluctuations induce electric fields in the Earth's surface via Faraday's law, generating low-frequency ($0.001 \to 0.1$ Hz) quasi-direct currents (GICs) that can flow into power grids \cite{oliveira_geomagnetically_2017,pulkkinen_geomagnetically_2017,piersanti_geoelectric_2019,ngwira_occurrence_2023}. 

When GICs flow over sensitive components within EHV transformers (e.g., windings), they disrupt normal operation by introducing nonlinearities in the magnetic core. This results in half-cycle saturation, which distorts the sinusoidal waveform, causes harmonic generation, and significantly increases magnetizing current demands \cite{lowery_voltage_2025,klauber_gic_2020,taran_effect_2023,clilverd_geomagnetically_2020,clilverd_geomagnetically_2025}. These distortions not only impair transformer efficiency but also accelerate insulation aging and thermal stress, reducing operational lifespan and potentially leading to permanent damage. Furthermore, GICs can impair the performance of grid protection systems: relays may misoperate, capacitor banks may overheat, and reactive power consumption may spike unexpectedly \cite{rajput_insight_2021,hughes_revealing_2022,heyns_geomagnetic_2021,wang_machine_2020}.

The magnitude and distribution of GICs are influenced by multiple factors, including the geospatial structure of the power grid, Earth's subsurface conductivity, and regional geomagnetic variability \cite{kataoka_extreme_2016,dimmock_gic_2019,kelbert_role_2020,gritsutenko_assessment_2023,sun_comparison_2019}. Coastal proximity and subsurface inhomogeneities can further amplify induced electric fields, heightening transformer exposure \cite{heyns_geomagnetic_2021}.

In extreme cases, these effects can culminate in widespread voltage collapse, cascading equipment failure, and large-scale blackouts \cite{dimmock_investigating_2024,hesami_naghshbandy_mitigating_2020}. The automatic disconnection of protective devices, while intended to prevent damage, may paradoxically trigger grid-wide instability and complicate recovery efforts such as black-start operations.

\subsection{The importance of quantifying space weather socio-economic impacts}\label{intro:economic-impact}

While numerous scientific studies have examined the physical characteristics of space weather and its effects on critical infrastructure, a notable lack of rigorous investigation remains into the broader socio-economic consequences that may arise from infrastructure failures induced by space weather events \cite{eastwood_economic_2017}. Indeed, business operations are vulnerable to space weather events due to potential disruptions in electricity supply, telecommunications, and other essential critical infrastructure services \cite{miteva_space_2023,oughton_economic_2021,xue_examining_2023,oughton_risk_2019}. A prominent example is the $1989$ geomagnetic storm, which caused substantial disruption to the Hydro-Québec electricity transmission network. The event activated protective systems designed to safeguard transformers, ultimately resulting in a large-scale blackout. The consequences included an estimated U.S. \$6.5 million in direct equipment damage, with total infrastructure repair costs reaching U.S. \$13.2 million \cite{bolduc_gic_2002, cander_ionospheric_2019}.

One global supply chain analysis suggests that a geomagnetic storm of similar magnitude to the 1989 Québec event could result in global GDP losses of up to U.S. $\$3.4$ trillion, approximately $5.6\%$ of the global economy \cite{schulte_how_2014}. These estimates are based on simplifying assumptions and do not incorporate direct coupling between space physics and engineering system models, meaning the results may overstate potential impacts. These scenario-based economic assessments have estimated the daily losses from extreme space weather events. \citeA{oughton_quantifying_2017} estimated U.S. daily losses ranging from \$6.2 billion to \$41.5 billion across four blackout-zone scenarios affecting between 8 and 66 percent of the U.S. population, with international supply-chain spillover adding a further \$0.8 to \$7 billion per day. They found that direct losses within the blackout zone averaged only 49 percent of the total cost. \citeA{eastwood_quantifying_2018} extended this approach to European exposures, parameterizing storm impacts by country-level grid resilience and forecast quality and propagating losses through value of lost load (VoLL) and multi-regional input--output models. For a two-substorm event over western Europe, they estimated direct losses of approximately \texteuro9.3 billion (U.S. \$10.2 billion) and international spillover of \texteuro787 to 1{,}108 billion (U.S. \$866 billion to \$1.2 trillion). By contrast, a quantitative macroeconomic assessment of potential space weather impacts on Japan estimated daily losses ranging from \textyen$19.8$ to \textyen$23.8$ billion (equivalent to U.S. $\$130-160$ million), arising from disruptions to electricity supply, navigation systems, and radio communications \cite{ishii_space_2021}. These huge differences in estimated impacts illustrate why we need better and more refined assessment methods. 

A widely used method for quantifying the macroeconomic impacts of critical infrastructure disruptions from natural hazards is input--output modeling. This approach is valuable for estimating cascading effects across national and regional economies, via supply chain disruption \cite{kelly_estimating_2015,galbusera_input-output_2018,sue_wing_economic_2020}. Some of the natural hazards assessed using these methods include hurricanes \cite{sprintson_assessing_2025}, floods \cite{lyu_evaluating_2023,jiang_assessing_2024,di_noia_high_2025}, rainstorms \cite{liu_quantifying_2023}, as well as space weather \cite{oughton_quantifying_2017}. These results can offer decision-makers actionable insights for risk mitigation and infrastructure resilience planning. Such insights are vital for informing decisions that promote the resilient design and operation of critical infrastructure systems. This section has emphasized that there is a strong need for new methodologies to help provide more reliable estimates of potential impacts.  

\subsection{A coupled physics--engineering--economic framework}

\begin{figure}
	\centering
	\includegraphics[width=1\textwidth]{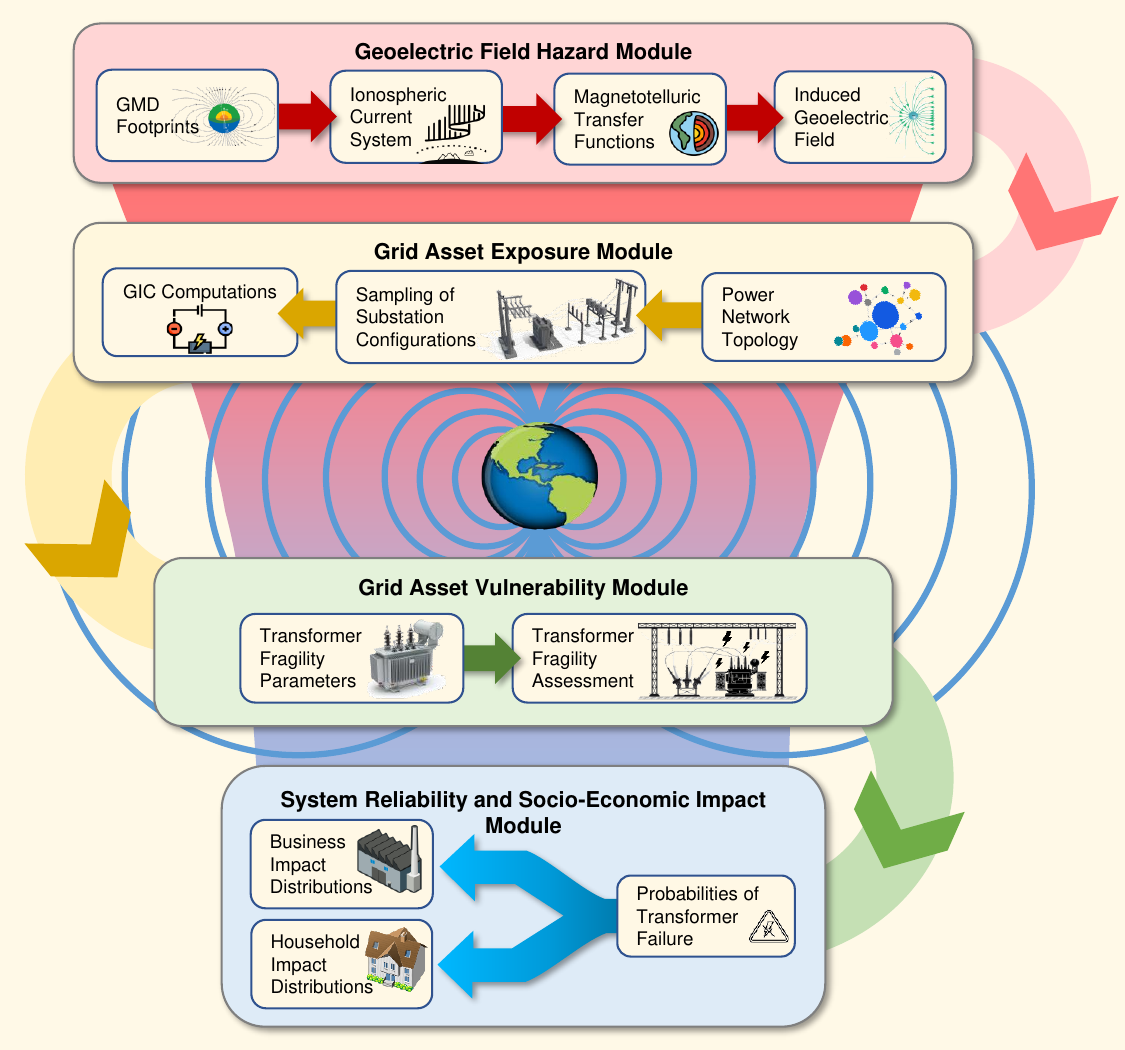} 
	\caption{\textbf{Graphical overview of the physics--engineering--economic coupling framework.} 
		The modules and data flows illustrated show how geophysical drivers are connected to socio-economic impacts through hazard characterization, engineering grid modeling, reliability assessment, and economic impact evaluation.}
	\label{fig:framework}
\end{figure}

We structure the assessment framework as a cascade from hazard to loss, as illustrated in Figure~\ref{fig:framework}. For example, the geoelectric field hazard module computes geoelectric fields from geophysical drivers using magnetic field interpolation and magnetotelluric transfer functions \cite{lucas_100-year_2020}, utilizing the NSF EarthScope USArray \cite{kelbert_methodology_2017, kelbert_em_2011}. The grid asset exposure module maps the geoelectric fields into equivalent voltage sources on transmission corridors and solves for GIC using a computationally efficient Lehtinen--Pirjola modified (LPm) formulation on a geospatial bulk power transmission network \cite{lehtinen_currents_1985, pirjola_lehtinen-pirjola_2022} (see Supporting Information Section 3 for implementation details).

The vulnerability module propagates effective per-phase GIC at transformers through age- and configuration-dependent fragility curves via Monte Carlo simulation to estimate substation failure probabilities and spatial patterns of grid disruption \cite{weimar_framework_2024}. The final socio-economic impact module builds service territories by Voronoi tessellation around bulk power substations, allocates population and industry activity to those territories \cite{thacker_geographic_2017}, and then maps localized service losses into direct and supply-chain-amplified economic impacts using an input--output model approach \cite{oughton_quantifying_2017}. The framework yields distributions of affected people, businesses, and losses for specified storm return periods, progressing the current state-of-the-art beyond deterministic estimates.

Specifically, the geoelectric field hazard modeling module derives storm-time surface magnetic fields from regional magnetometer networks using spherical elementary current systems (SECS) \cite{rigler_interpolating_2019} and translates fields into geoelectric vectors by convolving with frequency-domain magnetotelluric impedance tensors from national surveys \cite{love_empirical_2019}. Peak geoelectric amplitudes are fitted to lognormal distributions to obtain 100-, 150-, 200-, and 250-year return period scenarios, following the methodology of \citeA{lucas_100-year_2020} and \citeA{love_geoelectric_2018}. Geoelectric field vectors are integrated along transmission lines to produce segment-level electromotive forces, which, together with estimated line resistances, yield estimated GIC injections across the transmission network.

\begin{figure}
	\centering
	\includegraphics[width=1\textwidth]{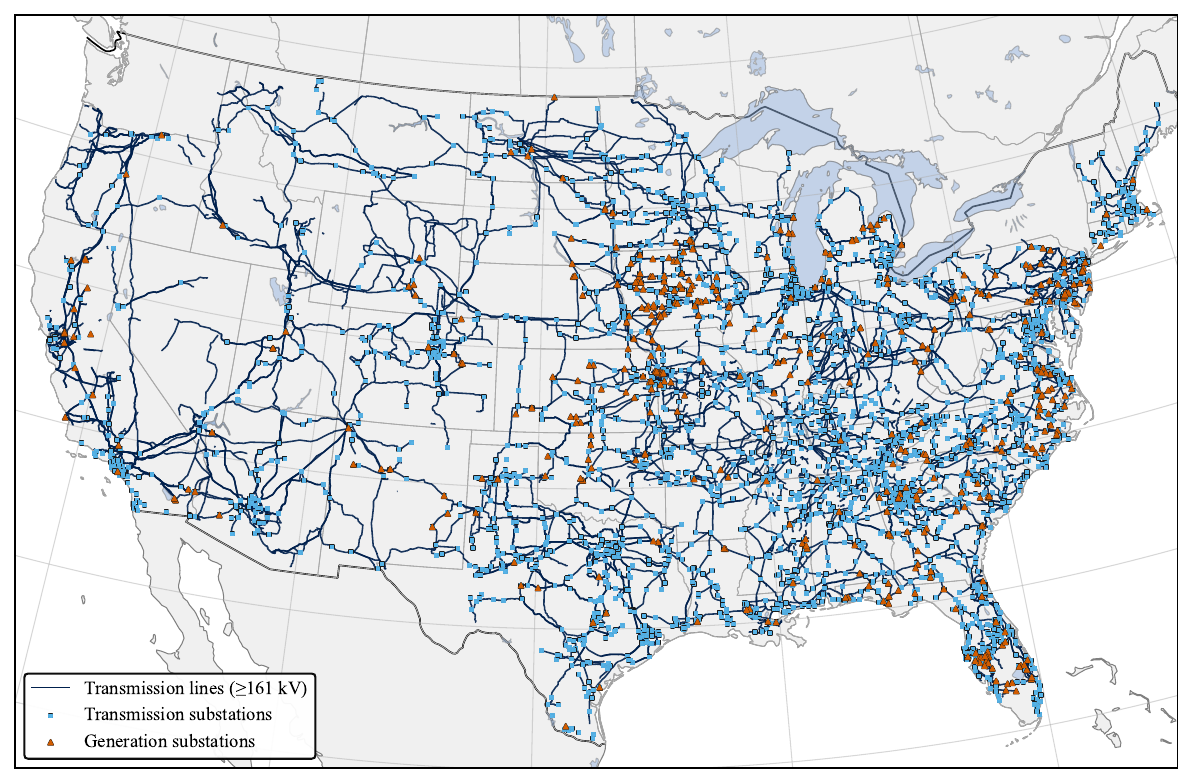} 
	\caption{\textbf{Geospatial bulk power transmission network model constructed from \textsc{OpenStreetMap} and HIFLD data.} 
		High-voltage transmission network ($\geq$161 kV) for the contiguous United States, comprising 10,464 substations and 16,256 transmission lines. This network is used in the risk modeling framework outlined in Figure~\ref{fig:framework}.}
	\label{fig:network}
\end{figure}

For the grid asset exposure module, we construct a contiguous U.S. bulk power transmission network, as illustrated in Figure~\ref{fig:network}, by intersecting OpenStreetMap substations \cite{openstreetmap_2024, raifer_overpass_2024} with Homeland Infrastructure Foundation-Level Data transmission lines \cite{hifld_transmission_2023} and assigning busbars by voltage class. We then represent lines, transformer windings, and grounds as a sparse DC admittance network and solve the LPm system to obtain nodal voltages, branch currents, and neutral-to-ground currents \cite{pirjola_lehtinen-pirjola_2022}. Effective per-phase GIC is computed from high-voltage winding and neutral currents scaled by nameplate voltages (via the turns ratio) \cite{klauber_gic_2020}. A Monte Carlo procedure samples transformer configurations (e.g., autotransformers, wye-wye, wye-delta), numbers of units per substation, grounding resistances, and parameter dispersion to reflect uncertainty and variability across the national transformer fleet.

Next, the grid asset vulnerability module obtains transformer failure probability modeled based on lognormal fragility, as a function of effective per-phase GIC. Hence, as the hazard metric of choice we utilize effective per-phase GIC because it better represents the actual thermal stress experienced by transformer windings compared to neutral current alone. While the neutral current $I_N$ is readily measurable, the effective per-phase GIC $I_{\text{E-GIC}}$ (see Supporting Information Eq.~S8) accounts for the current distribution through both high-voltage and low-voltage windings scaled by their voltage ratings, providing a more accurate indicator of hot-spot heating and the potential for protection relay misoperation that can disrupt supply. Dispersion reflects epistemic uncertainty and median capacity anchored to thermal-stress thresholds \cite{nerc_tpl-007-1_2014} (see Supporting Information Section 5 for full description). Age-related degradation is known to follow a Weibull distribution consistent with fleet demographics \cite{kabre_fragility_2022}. 

Finally, the system reliability and socio-economic impact module determines the number of failed substations. This information is converted into estimates of the affected population and businesses across service areas, and are used to compute direct economic losses by sector \cite{koks_understanding_2019}. We then apply the demand-driven input--output inverse to estimate total losses, capturing potential upstream supply chain impacts. To avoid arbitrary temporal restoration assumptions, daily loss distributions are reported.

\section{Materials and methods}

This is a summary of the materials and methods utilized, with the supplementary information providing a comprehensive overview. For the hazard characterization, we identify storm intervals from geomagnetic indices and interpolate surface magnetic fields using SECS. Geoelectric fields are derived in the frequency domain using magnetotelluric impedance tensors and transformed to time series, with peak statistics fitted to obtain return period scenarios. Further information can be found in the Supporting Information Section 1.

For the engineering grid characterization, we build a geospatial bulk power transmission network by intersecting substation and transmission datasets and assigning busbars by voltage. Lines and transformers are modeled as a DC admittance network. Further information can be found in the Supporting Information Section 2. Geoelectric voltages are converted to equivalent current sources, and the LPm system is solved to obtain nodal voltages and branch/neutral currents. Finally, effective per-phase GIC is computed from winding and neutral currents scaled by nameplate voltages. Further information can be found in the Supporting Information Section 3.

To evaluate the socio-economic impacts, Voronoi service areas are developed around the substation points to enable population and industry activity to be aggregated. County-level accounts are downscaled to daily, sectoral values using business establishment densities. Direct losses are propagated through a Leontief inverse to capture inter-industry linkages and summed to total daily losses. Further information can be found in the Supporting Information Section 4.

Taking a statistical approach to reliability, transformer failure probability is treated as following a lognormal fragility function with median capacity anchored to thermal-stress thresholds and dispersion sampled to reflect epistemic uncertainty. Age-related degradation follows a Weibull distribution derived from fleet statistics. Monte Carlo sampling is utilized over different transformer configurations, grounding, and parameter ranges, to yield many iterations of substation-level failure (with the true value logically expected to reside within the output distributions produced). Further information can be found in the Supporting Information Section 5.

To validate this approach, simulated GIC is compared to utility measurements for the recent severe ``Gannon'' storm from May 2024 using prediction efficiency, correlation, power spectral density, and magnitude-squared coherence. Interpolation residuals are propagated into the reliability module. Further information can be found in the Supporting Information Section 6.

\section{Results}

\begin{figure}
	\centering
	\includegraphics[width=1.0\textwidth]{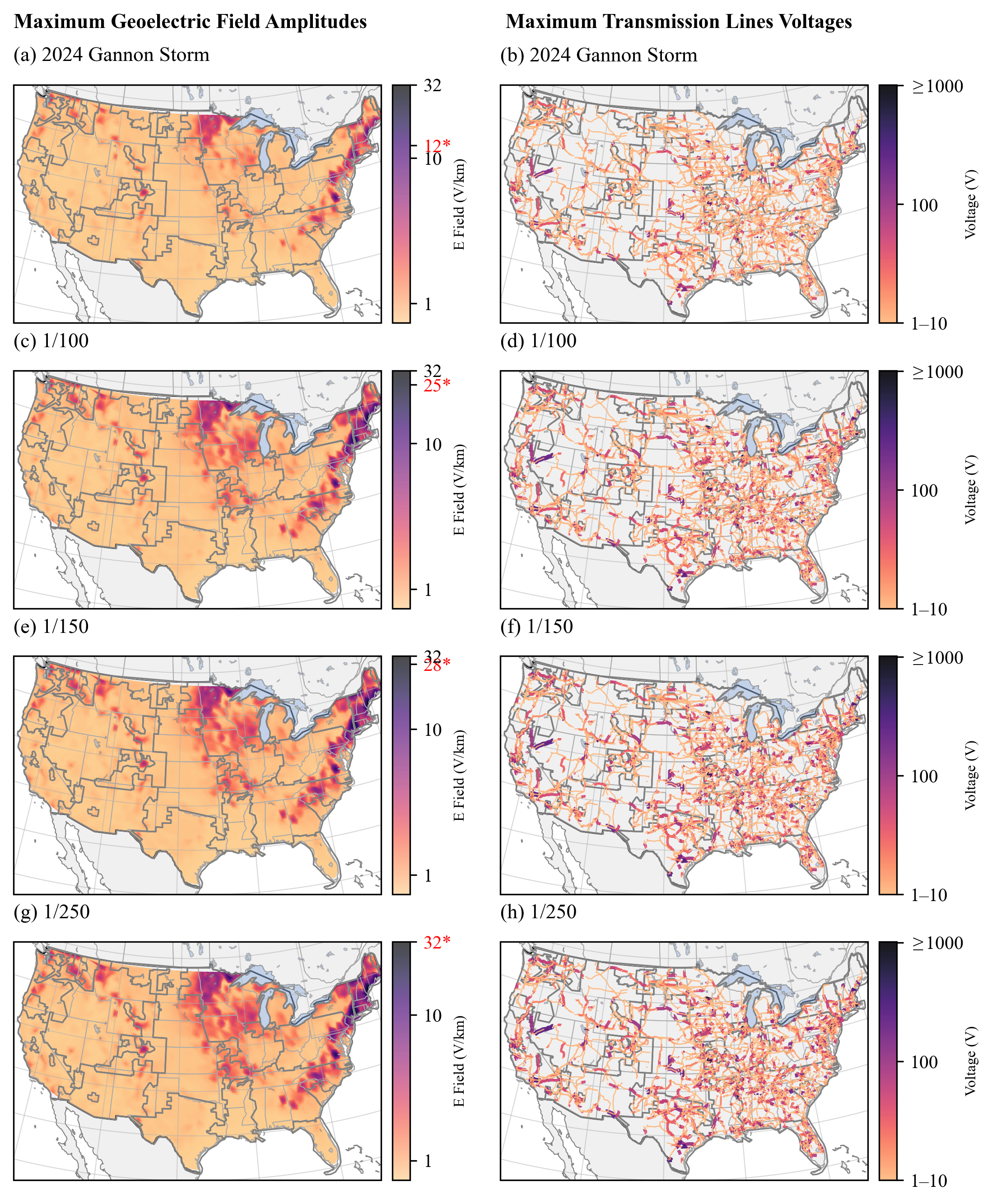} 
	\caption{\textbf{Extreme-value geoelectric field maps and induced transmission-line voltages for different return periods.} 
    Spatial distribution of peak geoelectric fields and corresponding induced voltages across the contiguous United States for 100-year, 150-year, and 250-year return period scenarios, with comparison to the recent severe ``Gannon'' storm for reference. Return period maps show statistical per-site maxima across the full storm catalog. The Gannon panels show a single spatially coherent snapshot (see Section 1 of Supplementary Information for peak selection methodology). The red value with an asterisk ($^*$) indicates the overall maximum across all locations within that panel.}
	\label{fig:geoelectric}
\end{figure}

The geoelectric field estimates illustrated in Figure~\ref{fig:geoelectric} show the largest response in the Great Lakes and northeastern seaboard areas. For 100-year, 150-year, and 250-year scenarios, peak fields reach $\sim$25, $\sim$28, and $\sim$32 V/km, respectively, with induced line voltages up to approximately $\sim$1.1 kV for the most severe cases. The recent severe ``Gannon'' storm exhibited isolated peaks near $\sim$12 V/km and comparable induced voltages, falling between the historical 50-year and 100-year activity levels. These patterns exhibit a strong spatial correlation with the underlying geoelectric field distribution, with induced line voltages further modulated by transmission line length, orientation relative to the geoelectric field vector, and path geometry. The geoelectric field patterns shown drive induced transmission-line voltages, which serve as the foundational hazard input ($H$ given site $S$) to the LPm power system model to estimate effective per-phase transformer GICs. For return period scenarios, these represent statistical estimates of peak amplitude at each site rather than a snapshot of a single physical moment, as described in Supporting Information Section 1. Moreover, this return period construction constitutes a conservative scenario in which the network is simultaneously exposed to the estimated peak amplitude at every site. Exposure risk is strongly concentrated in predominantly northern U.S. states, including the Great Lakes region, as well as running up the Appalachian mountains into the U.S. Northeast corridor.

\begin{figure}
	\centering
	\includegraphics[width=1.0\textwidth]{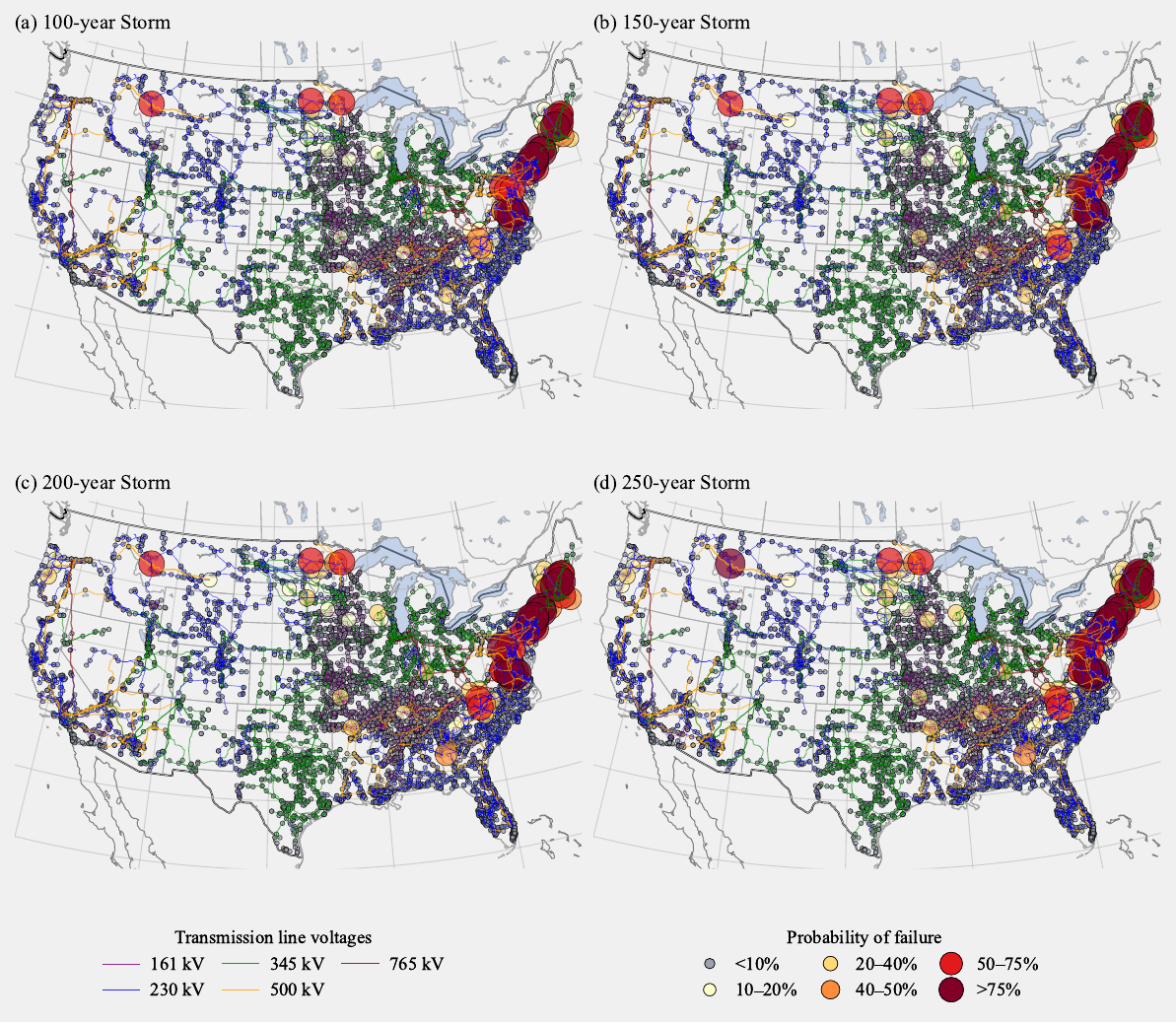} 
	\caption{\textbf{Estimated substation probability of failure under modeled GIC scenarios.} 
    Figures~a--d show the spatial distribution of transformer failure probability across the contiguous United States for different geomagnetic storm return periods. Failure probabilities are computed using the fragility-based reliability model described in Section 5 of the Supporting Information. Vulnerability hotspots (locations exceeding 50\% failure probability) are concentrated in regions with high ground conductivity and complex network topology, particularly in Wisconsin, Minnesota, the Upper Midwest, and coastal areas of the Northeast.}
	\label{fig:vulnerability}
\end{figure}

Using effective per-phase GIC as the hazard metric in fragility analysis, we find that the probability of transformer malfunction or protection system misoperation (our proxy for substation-level failure) rises systematically with storm severity. Indeed, this is illustrated in Figure~\ref{fig:vulnerability} with vulnerability highly localized by network topology and ground conductivity. Hotspots include Wisconsin, Minnesota, and the Upper Midwest, along with coastal concentrations in the Northeast. The Monte Carlo ensemble reveals that small changes in configuration, grounding, or asset age can significantly shift local failure probabilities, underscoring the importance of uncertainty quantification in risk communication.

\begin{figure}
	\centering
	\includegraphics[width=1\textwidth]{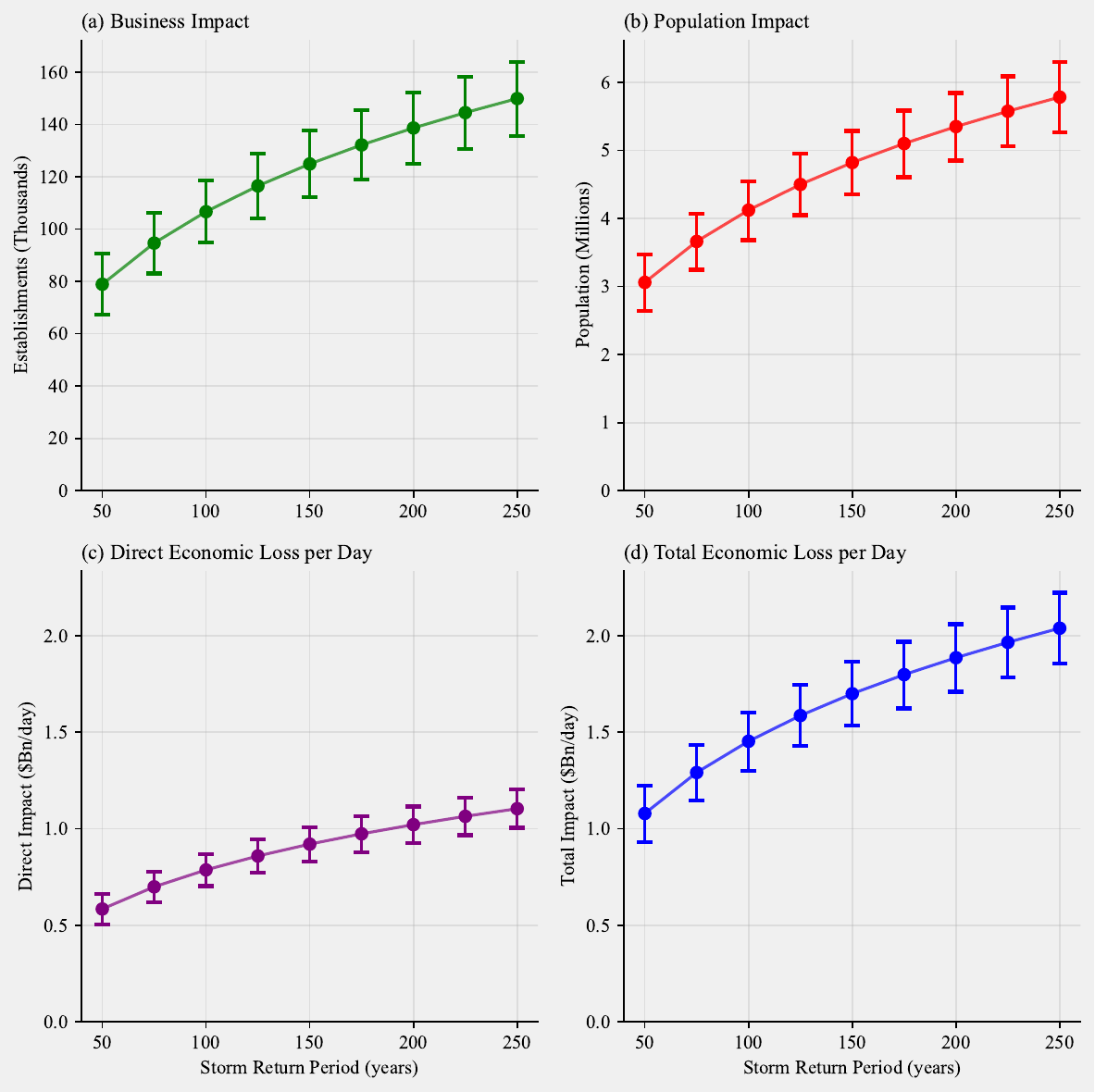} 
	\caption{\textbf{Socio-economic impacts versus return period: affected businesses, population, direct losses, and total losses.} 
		Results show systematic increases across all impact metrics with storm severity.}
	\label{fig:impacts}
\end{figure}

Aggregating across the ensemble to estimate socio-economic impacts, we find that a 100-year storm would disrupt power on the order of 4.1 million people and roughly 107,000 businesses on a given day, as illustrated in Figure~\ref{fig:impacts}. Subsequently, this could lead to direct economic losses of \$0.76 billion per day (\$0.64-0.88 at 95\%), and total losses, when accounting for inter-industry linkages, of up to \$1.45 billion per day (\$1.30-1.60 at 95\%). This contrasts with daily total economic loss estimates of \$1.70 billion for a 150-year storm (\$1.53-1.86 at 95\%), and \$1.89 billion for a 200-year storm (\$1.71-2.06 at 95\%). 

At 250-year intensity, affected populations rise toward six million and business disruptions toward approximately 150,000, with direct losses of roughly \$1.12 billion per day (\$0.97-1.28 at 95\%) and total losses near \$2.04 billion per day (\$1.86-2.22 at 95\%). The ratio of indirect to direct losses is broadly stable across severity, implying that shrinking the outage footprint or duration yields proportionate reductions in total losses.

\begin{figure}
	\centering
	\includegraphics[width=1\textwidth]{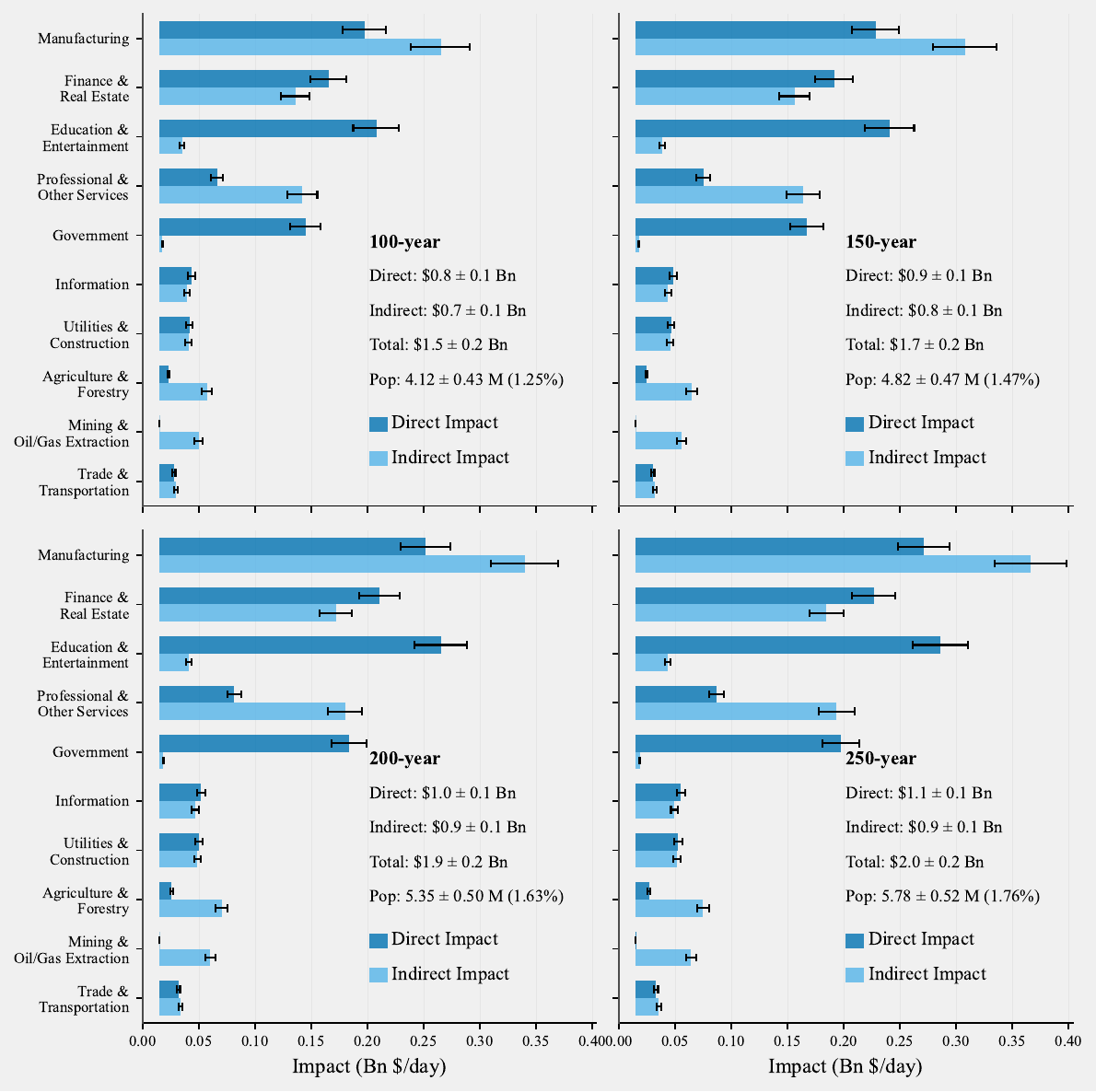} 
	\caption{\textbf{Sectoral breakdown of direct and indirect economic impacts for severe storm scenarios.} 
		Manufacturing, finance and real estate, and education and entertainment are the primary sectors experiencing losses in affected regions.}
	\label{fig:sectoral}
\end{figure}

As illustrated in Figure~\ref{fig:sectoral}, certain service sectors dominate daily losses in the most affected regions, particularly manufacturing, finance and real estate, and education and entertainment, followed by professional and other services and government. Trade and transportation, agriculture, and extractive industries are comparatively less exposed in the modeled events.

These results are generally a lower bound. The framework focuses exclusively on transformer thermal heating effects and excludes voltage collapse, cascading grid failures, restoration costs, and interdependent infrastructure disruption. The reported confidence intervals reflect parametric uncertainty in model inputs only and do not capture the full epistemic uncertainty associated with the storm scenario, future grid state, or societal exposure. The true societal risk is therefore likely substantially higher than the estimates presented here, and quantifying these additional dimensions remains a valuable direction for future research.

\section{Model validation}

We validate the end-to-end framework using data from the recent ``Gannon'' geomagnetic storm in May 2024, comparing simulated GIC at substations with utility measurements from Tennessee Valley Authority (TVA) \cite{wilkerson_gic--related_2025}. The model results are quantified using prediction efficiency and correlation in the time domain. Time-domain validation across multiple monitoring locations yields correlations ranging from $|r| = 0.02$ to $|r| = 0.70$ and prediction efficiencies from PE = -15.86 to PE = 0.48 (see Supporting Information Section 6.4). Then, in the frequency domain, we use power spectral density and magnitude-squared coherence to test correspondence across frequency bands that dominate transformer heating \cite{welch_use_1967, miranda_de_sa_note_2006}. Coherence is highest at low frequencies ($<0.004$ Hz), indicating that the hazard-to-GIC chain captures the slow variations most responsible for GIC flow, while short-period peaks are conservatively underestimated. This validation (as detailed in Supporting Information Section 6) justifies the use of the coupled framework for probabilistic assessment and communication of space weather risk in a field traditionally constrained by data availability.

\section{Global generalizability and transferability}

Although demonstrated using U.S. data, the framework is designed for transferability and to be used for national space weather assessments of other countries, with three necessary substitutions. First, replace magnetotelluric impedances and magnetometer networks with the best available national or regional conductivity and magnetic-field data. Where sparse, it is possible to use layered-Earth proxies or regionalized transfer functions calibrated to available stations. Second, substitute the national high-voltage grid with local transmission geodata, voltage classes, and substation inventories, which could potentially be drawn from national utilities, regional reliability coordinators, or curated open sources. Where data are incomplete, Monte Carlo sampling over plausible configurations and grounding resistances can bracket risk. Third, utilize local socio-economic statistical information, including disaggregated population estimates, business establishment microdata, and (national or multi-regional) input--output tables. These substitutions preserve the architecture of the hazard-to-loss cascade while anchoring the assessment in the local critical infrastructure stock and present economic structure.

The framework's modularity also supports comparative risk analysis. For example, countries with different geology, grid topology, and sectoral composition can be assessed on a consistent basis, enabling regional aggregation, cross-border policy coordination, and prioritization of investments to boost resilience. For emerging economies, where both grid expansion and vulnerability are evolving rapidly, the approach can provide early-stage guidance on substation siting, design, mitigation hardening, and operational preparedness. For highly interconnected economies, multi-regional input--output models allow supply-chain effects to cross borders, highlighting where shocks in one jurisdiction could propagate to others via trade in raw materials, manufactured goods, and digital services.

\section{Policy and operational implications}

End-to-end quantification of socio-economic risks improves how stakeholders develop response plans. For grid operators, spatial failure probability maps conditioned on storm severity guide the placement of GIC sensors, operational plans (such as grid reconfiguration and temporary load reductions), and selective hardening of substations with the highest value-at-risk. For regulators and planners, the framework could enable cost--benefit analyses that compare resilience options against expected reductions in daily losses (e.g., via blocking devices, enhanced grounding, forecasting and nowcasting, improved situational awareness, etc.). For the finance and insurance communities, the outputs provide a physics-grounded basis for pricing risk transfer, capital adequacy, and portfolio stress testing. For national policy makers, the framework provides a common language that connects geophysical drivers to households and economic impacts, facilitating cross-agency coordination and international cooperation on monitoring and mitigation. As the ratio of indirect to direct losses remains relatively stable within the modeled range, measures that reduce the outage footprint or speed restoration predictably translate into proportional reductions in total daily losses.

\section{Limitations and future work}

The present analysis is intentionally conservative in several respects. First, extreme-tail extrapolation in the hazard module is limited by the availability of historical data and the assumption that the maximum geoelectric field magnitude observed at each magnetotelluric site across all storms follows a lognormal distribution. As we continue to accumulate more magnetotelluric and storm event data, we will improve the return period calibration for space weather assessments such as this. Future research may also consider coupling geospace environment models to simulate many potential hazard scenarios, looking beyond only empirical events. Secondly, engineering assumptions are sampled to reflect plausible ranges but detailed asset registries would sharpen local estimates, such as transformer types, grounding values, line resistances, and the presence of blocking devices. Unfortunately, these proprietary data are not readily available for a national assessment of the selected country (the U.S.), and often fall under certain national security restrictions (such as controls on Critical Energy Infrastructure Information). Future research should look to undertake a detailed risk assessment with greater utility sector participation. Thirdly, the method does not account for network re-sequencing to mitigate GIC hotspot risk, as we unfortunately lack openly-available information to understand how this may take place. Fourthly, cascading power grid dynamics beyond initial substation-level failures are not modeled explicitly, such as voltage collapse, frequency disturbances, and interconnection-wide effects. Similarly, restoration duration is not modeled, as transformer replacement timelines range from days to over a year, depending on damage severity, supply chain availability, and grid reconfiguration. This means total event costs could substantially exceed the daily figures reported here. Prior studies suggest that small fractions of asset loss can trigger much larger system-wide disruptions, but the analysis needed to be bounded for tractability. Omitting these dynamics likely results in our economic estimates being a lower bound. Finally, we rely on aggregated economic data to simplify the modeling approach. Direct economic losses are estimated using a demand-driven Leontief input--output model, where the final demand shock is derived from the affected population fraction applied to Bureau of Economic Analysis (BEA) personal consumption and government expenditure data. A supply-driven Ghosh input--output formulation is retained in the supplementary information for comparison. Proportional allocation of value added to businesses within substation service zones introduces simplifying assumptions, and productivity heterogeneity within sectors and backup power for critical facilities (e.g., data centers, hospitals) are not modeled explicitly, which can both mitigate losses and add operating costs. The input--output structure is derived from BEA 2023 supply-use tables and 2020 Census population data, which may not fully capture the growing economic contribution of artificial intelligence and digital services --- sectors whose systemic importance and interdependencies are still emerging in national accounts. Future research should consider the global economic cascading effects of space weather disruptions, particularly as critical digital infrastructure, including data centers, cloud computing services, and artificial intelligence systems, becomes increasingly interconnected across national boundaries. Such analyses would require frameworks that extend beyond domestic input--output models to account for international supply chain dependencies and cross-border service disruptions, as well as dynamic models better suited to capturing long-run recovery effects.

\section{Conclusions}

We demonstrate a transferable, physics-grounded framework that links geophysical drivers to transformer stress and economy-wide losses, achieving the first coupled assessment of space weather impacts on infrastructure and the economy with quantified uncertainties. Using the U.S. as an example, we estimate daily economic losses for a 250-year geomagnetic storm from transformer thermal heating of \$2.04 billion (95 percent confidence interval: \$1.86 to \$2.22 billion), disrupting power for approximately 5.7 million people and 150,000 businesses. For a more frequent 100-year geomagnetic storm, we estimate daily economic losses of \$1.45 billion (95 percent confidence interval: \$1.30 to \$1.60 billion), with more than four million people and 107,000 businesses without power.

The findings provide justification for enhanced data collection, targeted monitoring, operational procedures, and selective hardening where value-at-risk is highest. Indeed, this evidence also offers a common template that countries can adapt to local conditions for space weather assessment (e.g., for different geologies, grid structures,  economic structures, etc.). Thus, by translating space weather risk into socio-economic impacts, the framework generates evidence for investments which enhance resilience, and supports national policy such as the U.S. National Space Weather Strategy and Action Plan \cite{sworm_national_2019}. Future work can begin to integrate cascading power grid dynamics and multi-hazard interactions (beyond the EHV transformer thermal heating focus addressed here), as well as model mitigation options.

\section*{Data Availability}
The C-SWIM analysis code and raw and processed data supporting this study are openly available in the C-SWIM GitHub repository (\url{https://github.com/denniesbor/C-SWIM}) and archived at Zenodo \cite{cswim_zenodo_2026}. Minute-resolution geomagnetic observatory data are obtained from INTERMAGNET (\url{https://imag-data.bgs.ac.uk/GIN/}), with underlying measurements provided by the U.S. \ Geological Survey Geomagnetism Program (\url{https://www.usgs.gov/programs/geomagnetism/data}) and Natural Resources Canada (\url{https://geomag.nrcan.gc.ca/data-donnee/sd-en.php}). Geomagnetic indices are from the Kyoto World Data Center for Geomagnetism (\url{https://wdc.kugi.kyoto-u.ac.jp/dstae/index.html}) and the GFZ German Research Centre for Geosciences \cite{matzka_geomagnetic_2021}. Magnetotelluric impedance tensors are from the EarthScope IRIS EMTF database \cite{kelbert_em_2011}.
The U.S. substation locations were sourced from OpenStreetMap (\url{https://www.openstreetmap.org}), and transmission line topology from the Homeland Infrastructure Foundation-Level Data transmission lines layer currently archived at (\url{https://arcgis.netl.doe.gov/server/rest/services/Hosted/Energy_Transition_Atlas_493d6/FeatureServer/18}). Demographic data are from the 2020 U.S. \ Decennial Census (\url{https://www.census.gov/library/visualizations/interactive/2020-population-and-housing-state-data.html}) and the 2023 ZIP Code Business Patterns (\url{https://www.census.gov/programs-surveys/cbp.html}), with ZIP to ZCTA concordance from the Health Resources and Services Administration (\url{https://data.hrsa.gov/GeoCareNavigator/Resources}). Economic input-output and regional GDP data are from the U.S. Bureau of Economic Analysis sector level use table (\url{https://www.bea.gov/industry/input-output-accounts-data}) and county GDP accounts (\url{https://www.bea.gov/data/economic-accounts/regional/}). Land cover data are from the USGS National Land Cover Database (\url{https://www.usgs.gov/centers/eros/science/national-land-cover-database}).
Measured geomagnetically induced current and geomagnetic field observations during the May 2024 Gannon storm are available from the SWERVE repository \cite{SWERVErepo2026, wilkerson_gic--related_2025}.

\section*{Conflict of Interest disclosure}
The authors declare no competing interests.

\acknowledgments
We acknowledge support from the U.S. National Science Foundation (NSF) through the NSF National Center for Atmospheric Research (via the Early-Career Faculty Innovator Program), which is a major facility sponsored by the U.S. National Science Foundation under Cooperative Agreement No. 1852977. Additional funding is recognized via the NSF RAPID ChronoStorm grant entitled ``Collecting Perishable Critical Infrastructure Operational Data for May 2024 Space Weather Events'' (No. 2434136). We thank collaborators in utilities and agencies for data access and validation discussions, especially the Tennessee Valley Authority.

\clearpage
\setcounter{figure}{0}
\setcounter{table}{0}
\renewcommand{\thefigure}{S\arabic{figure}}
\renewcommand{\thetable}{S\arabic{table}}

\clearpage
\appendix
\setcounter{figure}{0}
\setcounter{table}{0}
\renewcommand{\thefigure}{S\arabic{figure}}
\renewcommand{\thetable}{S\arabic{table}}

\appendix
\title{Supporting Information for ``Major Space Weather Risks Identified via Coupled Physics-Engineering-Economic Modeling''}

\authors{
Edward~J.~Oughton\affil{1},
Dennies~K.~Bor\affil{1},
Robert~S.~Weigel\affil{1},
C.~Trevor~Gaunt\affil{2},
Ridvan~Dogan\affil{1},
Liling~Huang\affil{1},
Jeffrey~J.~Love\affil{3},
and Michael~Wiltberger\affil{4}
}

\affiliation{1}{Space Weather Lab, George Mason University, Fairfax, VA, USA.}
\affiliation{2}{University of Cape Town, Cape Town, South Africa.}
\affiliation{3}{U.S. Geological Survey, Geologic Hazards Science Center, Denver, CO, USA.}
\affiliation{4}{U.S. National Science Foundation National Center for Atmospheric Research, Boulder, CO, USA.}


\newpage
\noindent\textbf{Introduction}

This section provides technical implementation details for the integrated framework. Data sources, software tools, parameter specifications, and detailed algorithmic procedures are described for each modeling component described in the main text.

\begin{figure}

\centering
\rotatebox{90}{%
    \includegraphics[width=0.95\textheight]{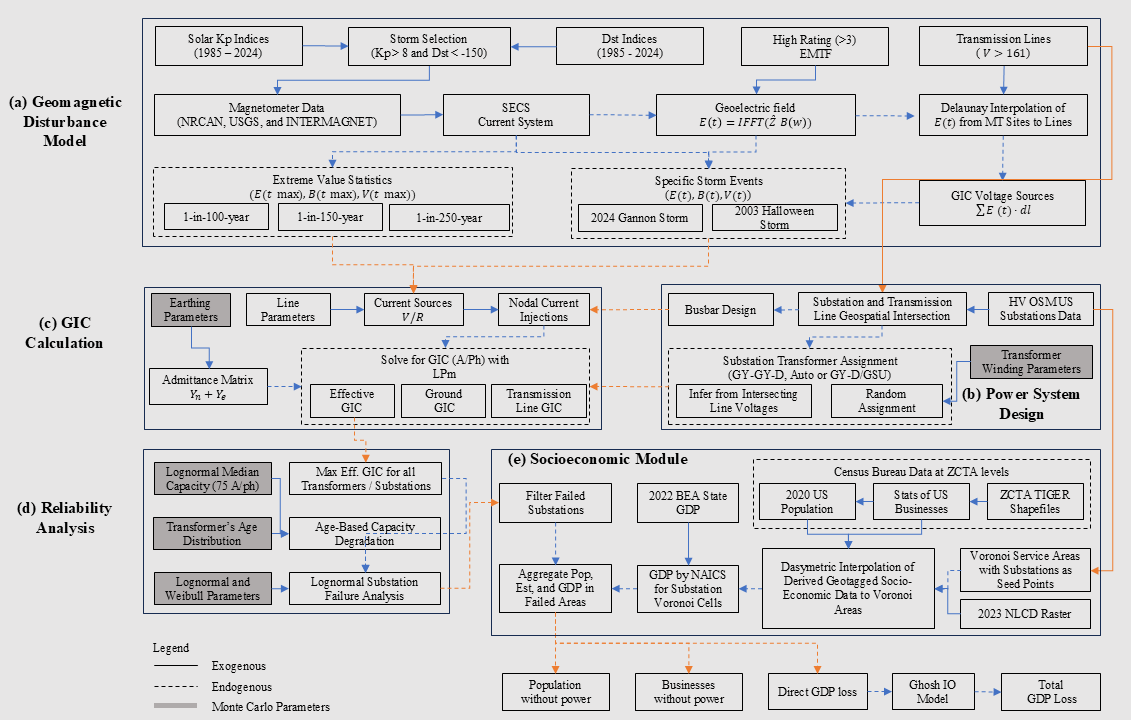}
}
\caption{\textbf{Detailed physics-engineering-economic method framework.} 
The figure illustrates the individual components of the risk and vulnerability assessment framework.}
\label{fig:S1}
\end{figure}

Fig.~\ref{fig:S1} illustrates the comprehensive framework, including data sources, software implementations, and technical parameters utilized in each modeling component. The framework integrates geomagnetic disturbance modeling, GIC calculation, and system bus admittance definition, power system network design, reliability analysis, and socio-economic impact assessment into a unified risk assessment pipeline. Exogenous data inputs (solid lines) and endogenous computational processes (dashed lines) flow through the five modules. Monte Carlo parameters (dark grey boxes) cascade uncertainty through each modeling step, enabling probabilistic risk assessment. Orange lines indicate information flows between major modeling components.

\section{Geomagnetic disturbance model}\label{method:gmd-footprint-app}

Data sources include the Dst index from the Kyoto World Data Center and the Kp index from GFZ Potsdam \cite{kyoto_plot_2024, matzka_geomagnetic_2021}. Geomagnetic storm periods are selected where Dst $<-140$~nT or Kp $>8$, following \cite{lucas_100-year_2020}. For each event, a 1.5-day buffer is applied before and after, and minute-resolution magnetic field data are obtained from INTERMAGNET \cite{intermagnet_2024}, USGS \cite{usgs_geomagnetism_2024}, and Natural Resources Canada \cite{nrcan_magnetic_2024}. Magnetotelluric (MT) impedance data are obtained from the NSF EarthScope USArray surveys \cite{kelbert_em_2011}, capturing regional variations in subsurface conductivity.

The horizontal geoelectric field is derived using the plane-wave approximation. In the frequency domain,
\begin{equation}
    \mathbf{E}(\omega,x,y) = \frac{1}{\mu_0}\,\mathbf{Z}(\omega,x,y)\,\mathbf{B}(\omega,x,y),
\label{eq:S1}
\end{equation}

\noindent
where $\mathbf{E}(\omega,x,y)$ is the horizontal geoelectric field (V/m), $\mathbf{Z}(\omega,x,y)$ is the $2 \times 2$ magnetotelluric impedance tensor ($\Omega$), $\mathbf{B}(\omega,x,y)$ is the horizontal magnetic field (nT), $\omega$ is angular frequency (rad/s), and $\mu_0 = 4\pi \times 10^{-7}$ H/m is the permeability of free space. Frequency-domain fields are transformed to the time domain via the inverse Fourier transform. 

Voltage sources (electromotive forces, EMFs) along transmission lines between nodes $n$ and $k$ are computed as
\begin{equation}
    V_{nk} = \int_{\ell_{nk}} \mathbf{E} \cdot d\mathbf{l}, \quad \text{(V)}
    \label{eq:S2}
\end{equation}

\noindent
where $d\mathbf{l}$ is the differential path vector (m) along transmission segment $(n,k)$. This is performed for all the derived bulk power transmission lines in the network.

Surface magnetic fields at MT sites are interpolated from magnetometer data using spherical elementary current systems (SECS) method \cite{rigler_interpolating_2019}, implemented in the PySECS Python package \cite{greg_lucas_2024_14511596}. SECS-interpolated fields were compared with independent magnetometer observations during the May 2024 ``Gannon'' storm to check the consistency of the interpolation used to drive geoelectric field estimates.

The computed geoelectric fields are interpolated from MT sites onto transmission line segments using Delaunay triangulation. To assess low-probability, high-impact geomagnetic disturbance events, lognormal distributions are fitted to historical peaks in $|E|$, $|V|$, and $|B|$ using the BezPy and Powerlaw Python libraries \cite{lucas_greglucasbezpy_2023, schaefer_schae234powerlaw_2017}. Return period fits are performed on absolute peak magnitudes of the SECS-interpolated magnetic field $\Delta B$ at MT sites, the MT-derived geoelectric field $E$, and the Delaunay-interpolated transmission line voltages $V$. These are extracted independently for each site and line across all storms in the 39-year catalog (1985--2024). Lognormal distributions are fitted via maximum likelihood estimation, with uncertainty bounds derived from bootstrap resampling (100 iterations). For scenario construction, per-line polarity is derived from the May~2024 (``Gannon'') storm by assigning the sign from near-peak samples (top $5\%$ of $|V|$). This is physically motivated by framing return period scenarios as extreme events whose spatial structure resembles that of the Gannon storm. The statistical fits yield absolute peak magnitudes, and the Gannon polarity provides the directional convention consistent with that spatial pattern. For 2 of the 16{,}256 transmission lines where the Gannon interpolation produces near-zero voltages at the network boundary, a conservative default polarity of $+1$ is assigned. For the Gannon storm, the peak snapshot is identified as the timestamp at which the median geoelectric field magnitude across all MT sites reaches its maximum, ensuring the maps in Fig. 3 represent a single physically coherent moment. This approach contrasts with the return period maps, where each site independently contributes its historical peak magnitude to the statistical fitting.

\begin{figure}
    \centering
    \includegraphics[height=0.8\textheight, keepaspectratio]{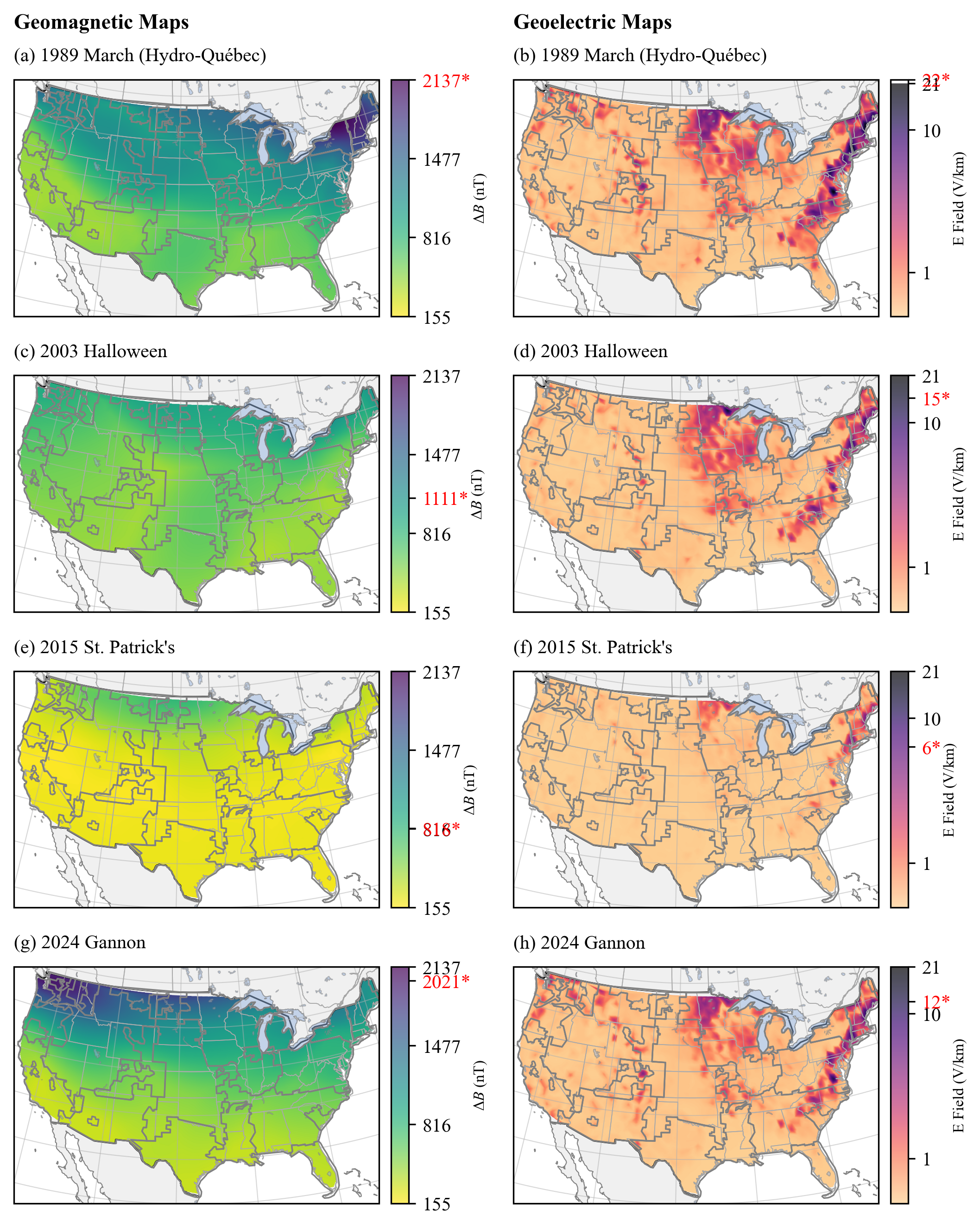}
    \caption{\textbf{Interpolated geomagnetic and geoelectric fields for historical storm events.} Left column shows magnetic field perturbations $\Delta B$ (nT), right column shows geoelectric field magnitude $E$ (V/km). Magnetic fields interpolated to MT sites using the SECS method, with a continuous grid generated via nearest-neighbor interpolation. (a-b) March 1989 (Hydro-Qu\'ebec), (c-d) October 2003 (Halloween), (e-f) March 2015 (St.\ Patrick's Day), (g-h) May 2024 (Gannon). Red asterisks indicate local spatial maxima.}
    \label{fig:historical_B_E_fields}
\end{figure}

\begin{figure}
    \centering
    \includegraphics[height=0.8\textheight, keepaspectratio]{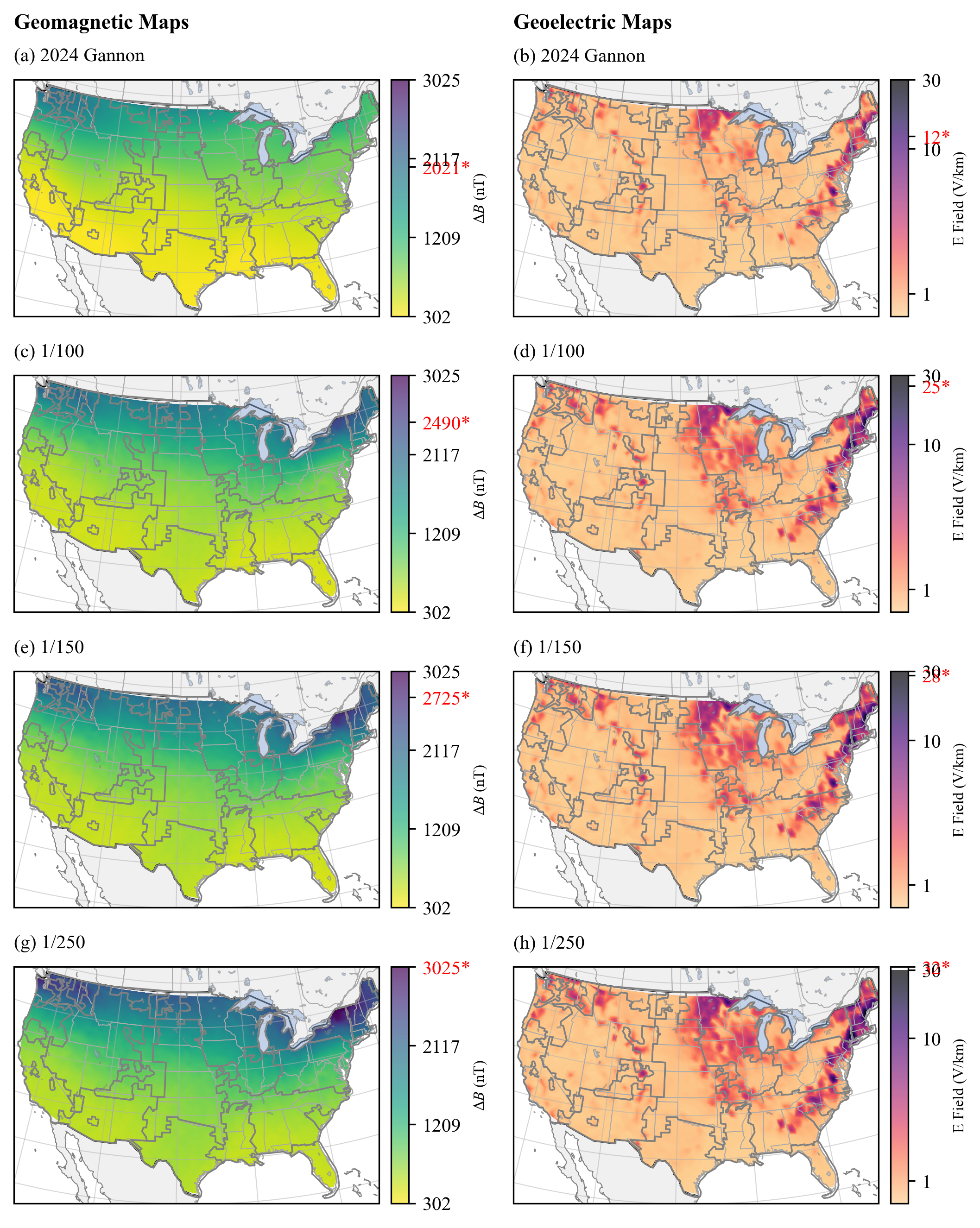}
    \caption{\textbf{Interpolated fields for extreme statistics scenarios.} Magnetic field perturbations $\Delta B$ (left) and geoelectric fields $E$ (right) for return period events derived from lognormal fits to historical peak magnitudes across the 39-year catalog (1985--2024). Fields interpolated to MT sites via SECS, then gridded using nearest-neighbor interpolation. (a-b) 2024 Gannon storm baseline, (c-d) 1-in-100 year event, (e-f) 1-in-150 year event, (g-h) 1-in-250 year event. Red asterisks indicate local spatial maxima.}
    \label{fig:extreme_B_E_fields}
\end{figure}

Figs.~\ref{fig:historical_B_E_fields} and \ref{fig:extreme_B_E_fields} show the interpolated magnetic and geoelectric field distributions for named historical storms and return period scenarios, respectively. The named storms are extracted from the 39-year geomagnetic storm catalog for comparison with the return period estimates.

\section{Power system network}\label{method:power-system-app}

This section outlines the creation of a geospatial nationwide power grid network for the contiguous United States. Location data for $\sim$60{,}000 substations were acquired from OpenStreetMap \cite{openstreetmap_2024, raifer_overpass_2024}, while transmission line data were sourced from the Homeland Infrastructure Foundation-Level Data (HIFLD) dataset \cite{hifld_transmission_2023}. These datasets were intersected to derive a geospatial power system graph where substations are nodes and transmission lines are edges. A 250-meter buffer was applied to each substation to improve intersection accuracy.  

The study focuses on high voltage and extra-high-voltage (EHV) networks at 161, 230, 345, 500, and 765~kV. Transmission lines at 115--138~kV are generally short, with higher resistance conductors, and thus contribute minimally to large-scale GIC flows. Hence, these lines were excluded.  

Substation busbars are defined by the voltage ratings of transmission lines connected to each substation. The highest voltage establishes the higher voltage (HV) bus, and the second-highest (if present) defines the lower voltage (LV) bus. The transformer type is then determined from these voltage pairs. Single-voltage substations are modeled as generator step-up units (GSU), with GY--D transformers, such that no GIC flows into the generator-connected side. For two-voltage substations, if the ratio $V_\text{HV}/V_\text{LV} < 2$, the transformer is assumed to be an autotransformer. If $V_\text{HV}/V_\text{LV} \geq 2$, the transformer is modeled as GY--GY, with both neutrals grounded. For three-voltage substations, the transformer is assigned as GY--GY--D, representing a typical three-winding configuration. Switchyards are identified where multiple transmission lines of the same voltage connect at a node, without other voltage levels or nearby generation.  

This deterministic methodology for busbar construction and transformer assignment has been applied in prior work \cite{oughton_reproducible_2024} and reviewed in detail during a technical workshop hosted at George Mason University (GMU) in September~2024, where the engineering assumptions were validated against industry practice.  

To capture uncertainties in transformer configurations, grounding parameters, and line resistances not specified by these deterministic assignment rules, a Monte Carlo approach is applied. Each substation is assigned between one and three transformers, depending on the type (one or two for GSU-like cases and up to three for transmission substations), with configurations sampled from standard U.S. designs, including GY-D, GY-GY, and Auto. Neutral grounding is explicitly modeled, with grounded-wye transformer neutrals providing return paths for GICs and substation grounding resistances sampled from a uniform distribution between 0.1~$\Omega$ or 0.2~$\Omega$. Equivalent DC winding resistances are assigned by transformer type (see Table~\ref{tab:power_system_params}), and line resistances are derived from HIFLD lengths (inflated by 3\% to account for sag and meander) with resistance per kilometer values assigned based on the line's voltage rating (Table~\ref{tab:power_system_params}).  

The resulting network is represented as a lumped-parameter DC admittance model, which includes line, transformer, and grounding impedances. For each of the 2,000 Monte Carlo simulations, nodal voltages are solved, and per-phase GIC flows are computed at each substation.

\section{GIC estimation with the Lehtinen--Pirjola modified method}\label{method:lpm-app}

Parameter values from the Horton benchmark model are used to construct the DC network admittance $\mathbf{Y}_n$ \cite{horton_test_2012}. These parameters are indicated in table \ref{tab:power_system_params}. Data-driven prediction of GIC at time $t$ is
\begin{equation}
    I_{\text{\small GIC}}(t) = a\,E_x(t) + b\,E_y(t),
    \label{eq:S3}
\end{equation}
\noindent
where $E_x$ and $E_y$ are the northward and eastward geoelectric components, and $a,b$ (A·m/V) depend on network topology \cite{ngwira_introduction_2019}. Nowcasting/forecasting with Equation~\ref{eq:S3} requires historical co-observations and an unchanged network. In this work, instead of relying on Equation~\ref{eq:S3}, we solve the network physics explicitly using the Lehtinen--Pirjola modified (LPm) method on the geospatially derived grid (Section~\ref{method:power-system-app}).

Relative to the classic LP formulation, LPm replaces the earthing impedance matrix $[\mathbf{Z}_e]$ with the earthing \emph{admittance} matrix $[\mathbf{Y}_e]$, sets ungrounded nodes to zero admittance, and solves directly for nodal voltages from the symmetric positive-definite system $(\mathbf{Y}_n+\mathbf{Y}_e)$ (efficiently via Cholesky decomposition), eliminating virtual grounds and enabling multi-voltage modeling. In the per-phase DC formulation adopted here, grounded substations receive diagonal entries $Y_{e,ii}=1/(3R_g)$ (per-phase), where $R_g\in\{0.1,0.2\}\,\Omega$ depends on the substation assignment. At the same time, ungrounded nodes have $Y_{e,ii}=0$.

Given the geoelectric voltage source $V_{nk}$ for each segment $(n,k)$ with resistance $R_{nk}$ and admittance $y_{nk}=1/R_{nk}$, an equivalent current source is
\begin{equation}
    j_{nk} = \frac{V_{nk}}{R_{nk}}.
    \label{eq:S4}
\end{equation}
This represents Norton's theorem applied to the transmission line segment, where the geoelectric field-induced voltage source is converted to an equivalent current source for circuit analysis. The current $j_{nk}$ flows as if the line segment were short-circuited, representing the driving current that would flow in the absence of other network impedances.

Current sources $j_{nk}$ are assembled into a current injection vector $\mathbf{J}_e \in \mathbb{R}^{N \times 1}$, and nodal voltages $\mathbf{V}_n \in \mathbb{R}^{N \times 1}$ solved from
\begin{equation}
    \mathbf{V}_n = \left( \mathbf{Y}_e + \mathbf{Y}_n \right)^{-1} \mathbf{J}_e,
    \label{eq:S5}
\end{equation}
\noindent
where $\mathbf{Y}_e \in \mathbb{R}^{N \times N}$ is the earthing admittance matrix, $\mathbf{Y}_n \in \mathbb{R}^{N \times N}$ is the network admittance matrix, and $N$ is the number of network nodes. The current flowing in branch $(n,k)$ is
\begin{equation}
    i_{nk} = j_{nk} + y_{nk}(v_n - v_k).
    \label{eq:S6}
\end{equation}
Substation earthing (neutral) currents are
\begin{equation}
    I_e = \mathbf{Y}_e \mathbf{V}_n,
    \label{eq:S7}
\end{equation}
equivalently $I_e=\mathbf{Z}_e^{-1}\mathbf{V}_n$ if $\mathbf{Z}_e=\mathbf{Y}_e^{-1}$ is used. The effective per-phase GIC is
\begin{equation}
    I_{\text{E-GIC}} = I_H + \left( \frac{I_N}{3} - I_H \right) \frac{V_X}{V_H},
    \label{eq:S8}
\end{equation}
\noindent
where $I_{\text{E-GIC}}$ is the effective per-phase GIC (A/ph), $I_H$ is the GIC in the high-voltage winding, $I_N$ is the neutral earthing current, and $V_H$, $V_X$ are the phase-to-neutral voltage ratings (kV) of the HV and LV windings, respectively.

The neutral earthing current $I_N$ (Eq.~\ref{eq:S7}) is the physical current flowing through the transformer neutral to ground, measured by GIC monitoring equipment at substations. The effective per-phase GIC $I_{\text{E-GIC}}$ (Eq.~\ref{eq:S8}) is a derived metric that accounts for current distribution through transformer windings and is used as the thermal stress hazard metric in fragility analysis. Moreover, it also represents the magnitude of currents that might cause protection relays to misoperate. For validation (Section~\ref{method:validation-gannon}), simulated neutral currents $I_N$ are compared with Tennessee Valley Authority (TVA) and North American Electric Reliability Corporation (NERC) measurements. For the vulnerability assessment (Section~\ref{method:reliability-math-app}), the effective per-phase GIC is employed, as it better represents thermal loading on transformer windings.

\section{Socio-economic impact assessment}\label{method:econ-analysis-app}

To quantify the societal consequences of transformer failures, we develop a spatial economic impact model that links power grid infrastructure to demographic and economic activity. Data sources include 2020 U.S. Census population data \cite{bureau_2020_2020}, 2023 ZIP-code-level business establishment data from the Census ZIP Code Business Patterns \cite{us_census_bureau_zip_2025}, 2023 county-level GDP data from the Bureau of Economic Analysis (BEA) Regional Economic Accounts (CAGDP2) \cite{bea_cagdp2_2025}. Both the establishment counts and county GDP contributions are organized by North American Industry Classification System (NAICS) sectors, allowing for sector-specific economic impact assessments. Business establishment data are mapped from ZIP codes to ZIP Code Tabulation Areas (ZCTAs) using 2020 concordance tables \cite{hrsa_zip_zcta_2024}. Final demand and inter-industry transaction data come from the 2023 BEA sector-level use table \cite{bea_use_2025}.

The service areas of each substation are modeled using a Voronoi tessellation of substation coordinates as seed points, which define service territories where each location is assigned to its nearest substation. County-level GDP is disaggregated to the ZCTA level by proportionally allocating economic output based on establishment density within each NAICS sector:
\begin{equation}
v_{s,z} = \frac{N_{s,z}}{\sum_{j \in c(z)} N_{s,j}} \cdot G_{s,c(z)}
\label{eq:S9}
\end{equation}
\noindent
where $v_{s,z}$ is the estimated GDP (\$/day) for sector $s$ in ZCTA $z$, $c(z)$ denotes the county containing ZCTA $z$, $N_{s,z}$ is the number of NAICS sector $s$ establishments within ZCTA $z$, $\sum_{j \in c(z)} N_{s,j}$ is the total establishments in sector $s$ across all ZCTAs in county $c(z)$, and $G_{s,c(z)}$ is the daily GDP of sector $s$ in that county. This approach assumes economic activity scales with establishment density, though it may not capture variations in establishment size or productivity within sectors. Annual county GDP values are converted to daily rates by dividing by 365 to estimate daily economic losses during power outages.

Socio-economic data are spatially redistributed from ZCTAs to substation service areas using masked dasymetric interpolation \cite{eicher_dasymetric_2001}. The 2023 National Land Cover Dataset (NLCD) developed areas (classes 21-24) provide spatial weights to concentrate economic activity in built environments rather than assuming uniform distribution \cite{usgs_national_2020}. Following spatial redistribution, $P_i$ represents the population within substation service area $i$. Economic activity is aggregated into ten major sectors: agriculture, mining, utilities and construction, manufacturing, trade and transportation, information, finance and real estate, professional and other services, education and entertainment, and government, where $v_{s,i}$ represents the GDP (\$/day) for sector $s$ within substation service area.

The total affected population from transformer failures is calculated as:
\begin{equation}
L_{pop} = \sum_{i \in F} P_i
\label{eq:S10}
\end{equation}
\noindent
where $F$ represents the set of failed substations and $L_{pop}$ is the total affected population. The total direct sectoral losses are calculated by summing affected economic activity within failed service territories:
\begin{equation}
v_s = \sum_{i \in F} v_{s,i}
\label{eq:S11}
\end{equation}
\noindent
where $v_s$ is the total direct loss (\$/day) in sector $s$.

To capture inter-sectoral dependencies, $v_s$ and $L_{pop}$ are propagated through the wider economy using an input--output (IO) framework. Two shock models are considered. The first is a Leontief demand-driven model \cite{miller_input-output_2009}, which treats the outage as a reduction in final demand and propagates it backward through inter-industry demand. The second is a Ghosh supply-driven model \cite{ghosh_input-output_1958}, which treats $v_s$ as a reduction in sector value added and propagates it forward through the supply chain. We adopt the Leontief model as the primary model because the physical trigger of a power outage is the withdrawal of electricity from end users, which directly prevents final consumption.

The daily final demand shock is constructed from the affected population fraction $\rho = L_{pop} / N^{US}$, where $N^{US}$ is the 2020 decennial U.S. population:
\begin{equation}
\Delta \mathbf{f} = -\frac{\rho}{365}\left(\mathbf{f}^{PC} + \mathbf{f}^{GOV}\right),
\label{eq:S12}
\end{equation}
\noindent
where $\mathbf{f}^{PC}, \mathbf{f}^{GOV} \in \mathbb{R}^{10}$ are the personal consumption and government consumption columns of the BEA use table aggregated to the ten sectors used throughout this study. Government consumption is shocked in proportion to population because most government services are delivered to people, and population is the natural proxy for services reachable during a regional outage. The total daily economic impact is computed through the Leontief inverse,
\begin{equation}
\Delta \mathbf{x} = \mathbf{L}\,\Delta \mathbf{f}, \qquad \mathbf{L} = (\mathbf{I} - \mathbf{A})^{-1},
\label{eq:S13}
\end{equation}
\noindent
where $\mathbf{A} \in \mathbb{R}^{10 \times 10}$ is the direct requirements matrix derived from the BEA use table.

The Ghosh model propagates direct sectoral losses through the Ghosh inverse,
\begin{equation}
\mathbf{x}' = \mathbf{v}'\mathbf{G},
\label{eq:S14}
\end{equation}
\noindent
where $\mathbf{G} \in \mathbb{R}^{10 \times 10}$ is the Ghosh inverse and $\mathbf{v}'$ is the row vector of direct sectoral losses. The Ghosh model has been criticized on theoretical grounds \cite{oosterhaven_plausibility_1988} for implying that downstream sectors absorb upstream output independently of price or input substitutability, which does not hold under most realistic disruption scenarios. It is therefore retained for comparison, and the results are reported in Section~\ref{sec:ghosh-econ-model}.

\section{Reliability analysis}\label{method:reliability-math-app}

Here, a methodology is defined to estimate substation failure probability based on GIC exposure. To assess cascading impacts from space weather events to societal losses, we employ a value-at-risk (VaR) framework following a cascading probability structure:
\begin{equation}
\text{VaR} = P(H|S)P(R|H)P(D|R)P(L|D) 
\label{eq:S15}
\end{equation}
\noindent
where $S$ is the critical-infrastructure site, $H$ the hazard (space-weather footprint), $R$ the system response, $D$ the component-damage state, and $L$ the resulting loss \cite{weimar_framework_2024}. Having modeled the hazard footprint (Section~\ref{method:gmd-footprint-app}) and characterized the infrastructure network (Section~\ref{method:lpm-app}), we now quantify system response through fragility curves that describe failure probabilities and associated losses, using Monte Carlo simulation to capture uncertainties in component behavior.

Electrical component failures follow lognormal fragility curves \cite{kabre_fragility_2022}, therefore, we model transformer failure probability given GIC exposure as:
\begin{equation}
P_{\text{fail}} = \Phi\left[\frac{\ln I_{\text{E-GIC}} - \ln \theta_0}{\beta}\right]
\label{eq:S16}
\end{equation}
\noindent
where $\Phi$ is the cumulative distribution function of the standard normal distribution, $\theta_0$ is the GIC median capacity, and $\beta$ is the lognormal dispersion parameter. TPL-007 guidelines specify 75 A/ph as the threshold requiring thermal stress analysis \cite{nerc_tpl-007-4_2020}, and we adopt this as $\theta_0$, representing 50\% probability of transformer maloperation that may trigger system-wide failure.

Due to the limited availability of transformer-specific dispersion data, the lognormal standard deviation is treated as epistemic uncertainty, sampled as $\beta \sim \mathcal{U}(0.25, 0.50)$ within the recommended interval for electrical components \cite{fema_multi-hazard_2001,iaea_approaches_2019}. Age-related degradation is incorporated through a two-parameter Weibull distribution:
\begin{equation}
F_{\text{age}}(a) = 1 - \exp\left[-\left(\frac{a}{\eta_{\text{age}}}\right)^{\beta_{\text{age}}}\right] 
\label{eq:S17}
\end{equation}
\noindent
where $a$ is transformer age, and scale and shape parameters are sampled as $\eta_{\text{age}} \sim \mathcal{U}(30, 50)$ years and $\beta_{\text{age}} \sim \mathcal{U}(1, 3)$. Based on fleet statistics showing 55\% of U.S. distribution transformers exceed 33 years \cite{mckenna_distribution_2024}, and assuming this age distribution is representative of the broader transformer fleet, including high voltage and EHV transmission units, we sample 55\% of ages from [33, 50] years and 45\% from [1, 32] years. The baseline failure threshold is reduced by this age factor (up to 40\% capacity loss). Simulated GIC values are adjusted by a factor uniformly drawn between 0.6 and 1.4 to account for model prediction uncertainties (Fig.~\ref{fig:gic_time_comparison}).

The Monte Carlo simulation samples fragility parameters and age distributions independently for each iteration to capture epistemic uncertainties in transformer failure behavior. For each simulation iteration, transformer failure states are determined using Bernoulli trials based on the computed failure probabilities. When a substation fails, the affected population and economic activity within its Voronoi service territory are aggregated using Equations~\ref{eq:S10} and \ref{eq:S11} to quantify direct impacts, with total economic losses computed using Equation~\ref{eq:S13}.

The simulation continues until the relative half-width of the 95\% confidence interval for mean population loss satisfies $h_n / \hat{\mu} < 0.05$, where $\hat{\mu}$ and $h_n$ are the sample mean and confidence interval half-width, ensuring a maximum 5\% relative error. Each simulation iteration yields one realization of potential demographic and economic impacts, forming an ensemble that provides a probabilistic risk profile for the space weather event.

\begin{figure}
    \centering
    \includegraphics[width=1\linewidth]{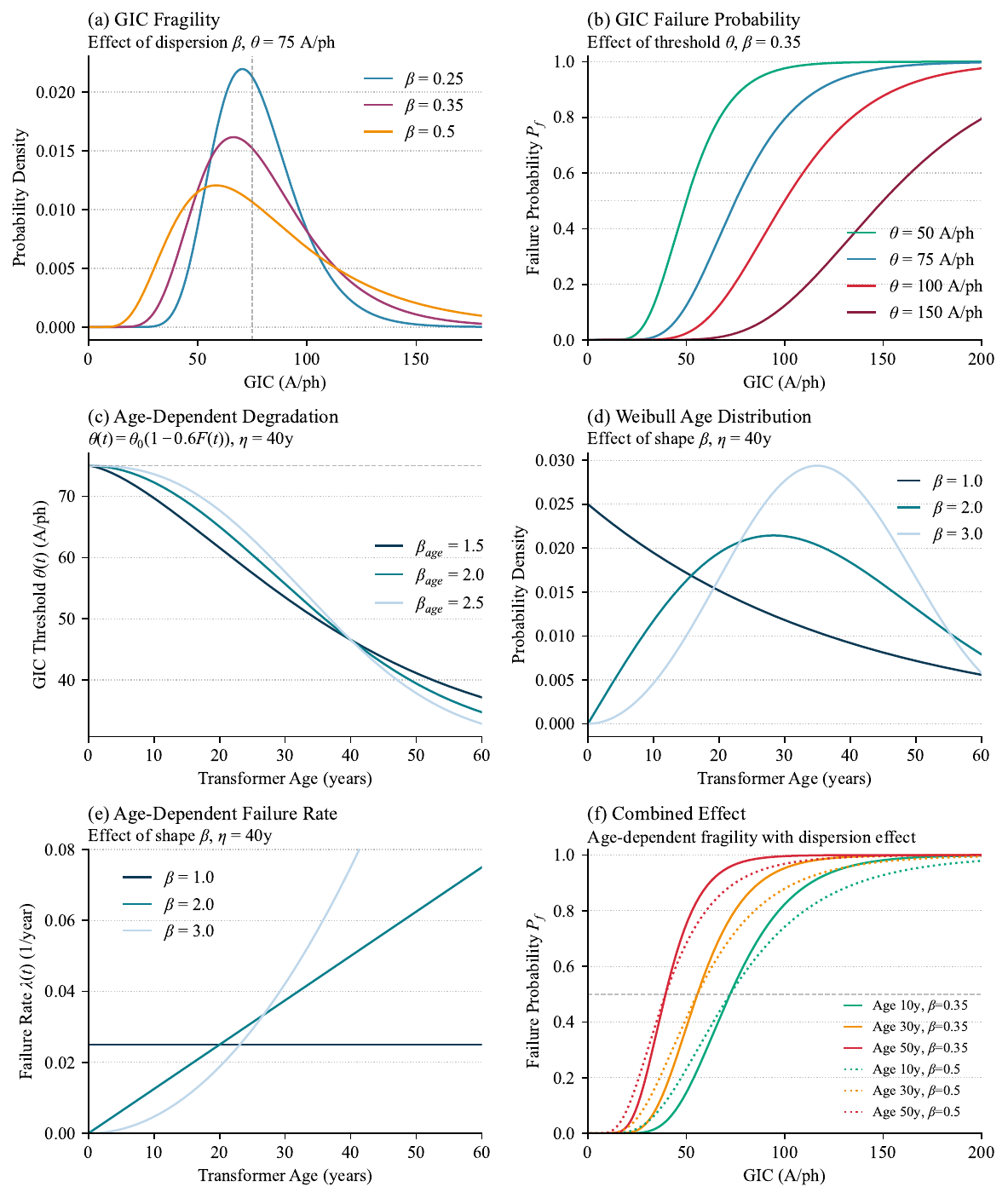}
    \caption{\textbf{Fragility curves for transformer failure under GIC exposure.} Effects of (a) lognormal dispersion $\beta$, (b) threshold capacity $\theta$, (c) age-dependent degradation, (d-e) Weibull aging parameters, and (f) combined age and dispersion on failure probability.}
    \label{fig:reliability_analysis}
\end{figure}

Monte Carlo simulation parameters are summarized in Table~\ref{tab:reliability_params}, with representative fragility curves illustrating the effects of threshold capacity, lognormal dispersion, and age-dependent degradation shown in Fig.~\ref{fig:reliability_analysis}.

\section{Model Validation and Coupling Assessment}\label{sec:validation-coupling}

This section validates the complete GIC modeling chain from external drivers to network response. We test power-system sensitivity on a benchmark grid, verify event-scale behavior against measurements during the May 2024  ``Gannon'' storm, quantify errors introduced by magnetic-field interpolation, and compare a data-driven alpha--beta predictor with physics-based simulations across return period scenarios. The goal is to assess each coupling step and its aggregate impact on predicted GIC.

\subsection{Physics-based model sensitivity -- validation on the Horton benchmark grid}\label{method:horton}

Similar to our stochastic approach, we first isolate power-system physics by imposing controlled electric-field forcing on a standard test network \cite{horton_test_2012} to quantify sensitivity to grounding, transformer configuration, and GIC blocking. This test grid is herein referred to as the Horton grid.

\begin{figure}
    \centering
\includegraphics[width=1\linewidth]{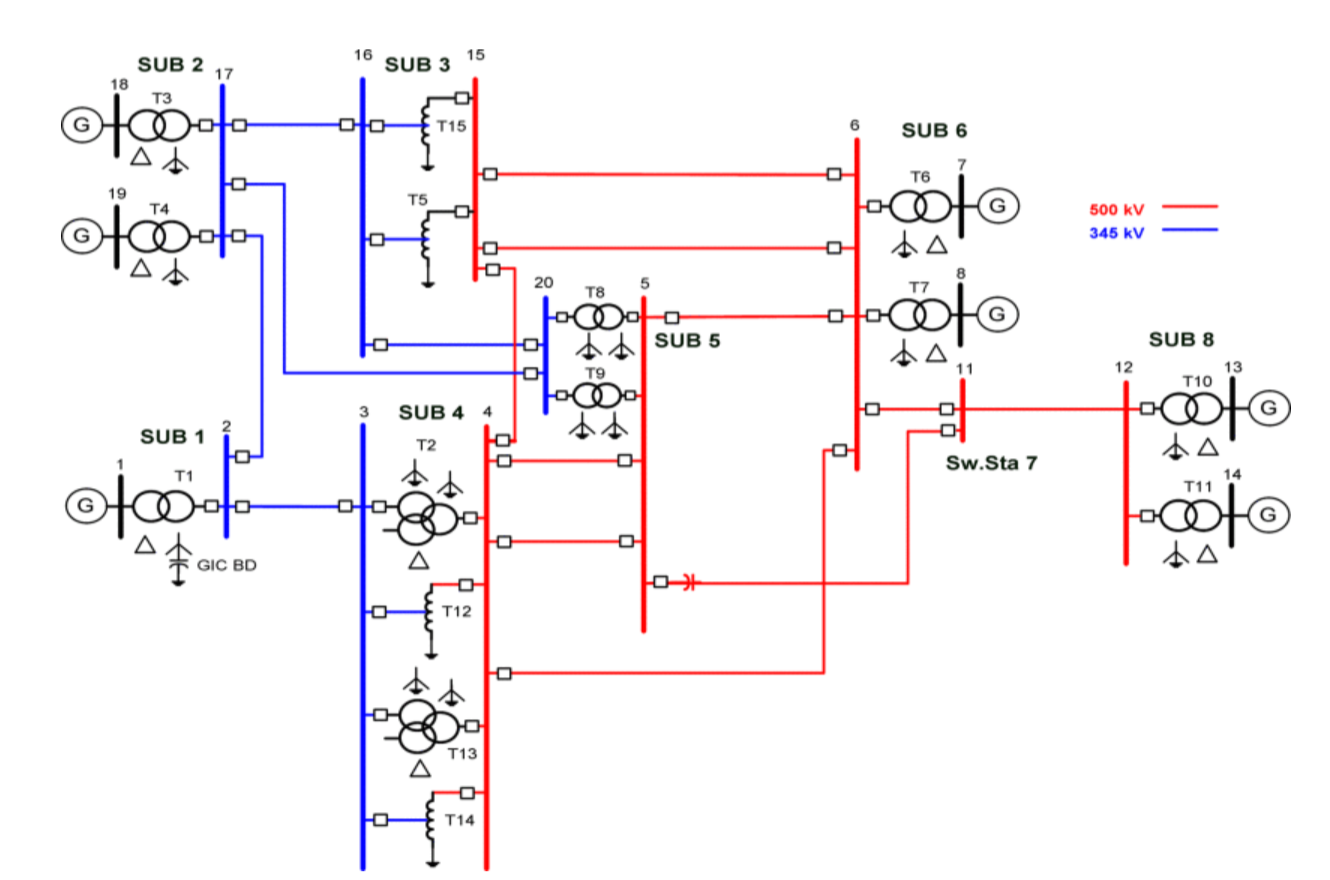}
    \caption{\textbf{Horton benchmark test system.} Test system configuration adapted from \cite{horton_test_2012}.}
    \label{fig:horton-grid}
\end{figure}

The Horton grid is a 20-bus power system consisting of eight substations with diverse transformer and transmission line architectures. In the benchmark specification, a single generator step-up (GSU) transformer is equipped with a GIC blocking device, one transmission corridor is blocked, and a single substation functions as a switchyard. Fig.~\ref{fig:horton-grid} illustrates the Horton configuration, adapted from the original reference.

The base Horton model was reproduced deterministically to ensure consistency with published results. We then generated stochastic variants of the Horton network to capture structural uncertainty. Transformer configurations were perturbed among GY--GY, Auto, and GY--GY--D types, and neutral grounding states were randomized. Generator step-up units could be toggled between blocked and unblocked states, while non-GSU substations were probabilistically assigned infinite grounding resistance to emulate blocking. Substation grounding resistances were otherwise drawn from a uniform distribution on $[0.1,0.2]~\Omega$. Transmission lines were additionally subject to probabilistic GIC blocking.

A uniform $1~\text{V/km}$ geoelectric field was applied separately in the northward and eastward directions to provide controlled forcing. For each Monte Carlo simulation, the Lehtinen--Pirjola modified (LPm) method (Section~\ref{method:lpm-app}) was solved to obtain nodal voltages and currents. In total, 5,000 simulations were performed. From these, substation-level ground currents, line flows, and transformer winding GICs were computed. The results quantify the sensitivity of GIC exposure to uncertainties in grounding, transformer configuration, and line availability, providing a distributional rather than deterministic estimate of GIC magnitudes.

\begin{figure}
    \centering
    \includegraphics[width=1\linewidth]{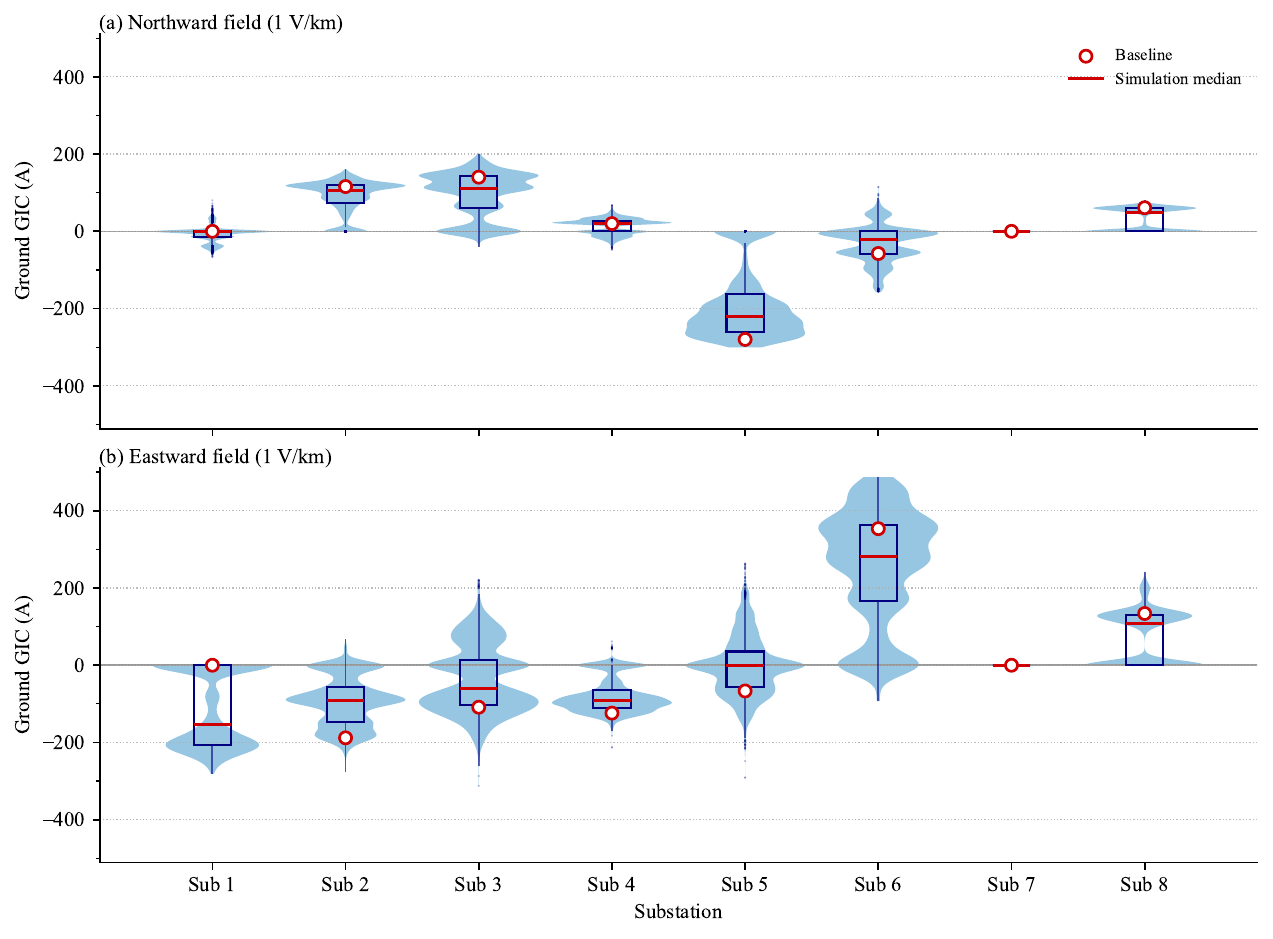}
    \caption{\textbf{Distribution of ground GICs in the Horton benchmark grid under a $1~\text{V/km}$ field.} Monte Carlo simulations (violin plots) compared with the deterministic baseline (red circle markers).}
    \label{fig:horton-results}
\end{figure}

Fig.~\ref{fig:horton-results}(a)--(b) shows ground GIC distributions for the eight substations in Fig.~\ref{fig:horton-grid} under uniform $1~\text{V/km}$ northward and eastward forcing. Across 5,000 Monte Carlo draws, the baseline (red circle) aligns closely with the sample median (red square) at most sites, indicating near-symmetric response over the tested parameter ranges. Field orientation redistributes exposure. The northward case concentrates currents at Substations 2--4 (peaks $\sim10^2$~A) and produces a pronounced negative tail at Substation 5. The eastward case shifts the most significant currents to Substation 6 (a few $\times10^2$~A). Variance is driven by grounding state, transformer type, transmission line, and GSU blocking configurations. When a substation is effectively ungrounded, the ground current collapses toward zero. Substation~1 is a GSU (D$\to$gen--GY) with a blocking device in the benchmark: when blocked, ground GIC is suppressed locally, and unblocking introduces nonzero currents that redistribute flows across the network. The deterministic baseline lies within the stochastic distributions at all sites, confirming replication of the benchmark and quantifying spread due to plausible network uncertainty. These controlled experiments validate our methodology before applying the same LPm-based workflow to assess substation GIC exposure across the U.S. using open-source network, MT, and magnetometer datasets.

\subsection{Event-scale validation against TVA measurements (May~2024)}\label{method:validation-gannon}

We next validate the end-to-end behavior by comparing the simulated substation GIC with independent TVA and NERC measurements from 10--12 May 2024. This links magnetic drivers, MT-based impedances, and LPm network response to observed currents. TVA and NERC GIC monitoring instruments measure neutral-to-ground current at substation transformers. Thus, we compare measured values with simulated neutral earthing currents $I_e$ from Eq.~\ref{eq:S7}. The dataset was acquired from \cite{wilkerson_gic--related_2025}. We use a Haversine distance algorithm to determine the nearest substations to GIC monitor locations \cite{maria_measure_2020}, with spatial metadata in EPSG:4326 transformed to EPSG:3857. TVA reports coordinates to four decimal places, corresponding to roughly 11~m precision in latitude, while NERC measurements are reported to zero decimal places, limiting spatial accuracy to the kilometer scale. TVA monitor locations are matched to OSM substations within a 100-meter radius, treating each monitor as connected to the substation's neutral grounding. NERC measurements are evaluated using the same procedure.

\begin{figure}
    \centering
    \includegraphics[width=1\linewidth]{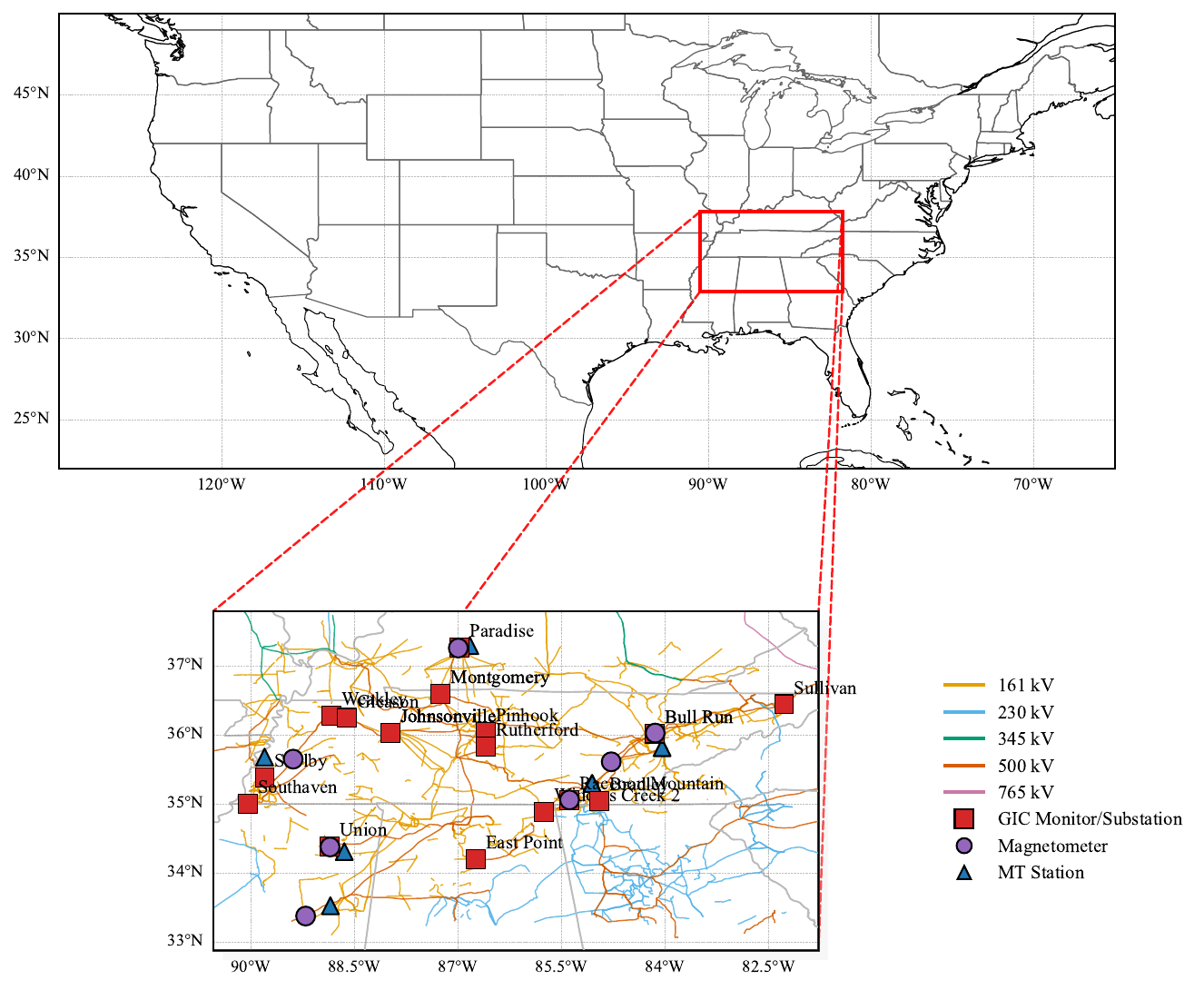}
    \caption{\textbf{TVA validation infrastructure.} TVA GIC monitors matched with co-located OSM substations within 100~m, bulk power transmission lines (161, 230, 345, and 500~kV), magnetometer stations, and MT sites used for model validation during 10--12 May 2024.}
    \label{fig:tva_sites}
\end{figure}

Fig.~\ref{fig:tva_sites} indicates the TVA validation infrastructure, including GIC monitors, co-located substations, transmission lines, and measurement sites. The analyzed storm window spans 10 May 12:00~UTC to 12 May 12:00~UTC.

Time-domain validation employs prediction efficiency (PE) and the Pearson correlation coefficient, where PE is defined as:
\begin{equation}
\label{eq:S18}
\mathrm{PE} = 1 - \frac{\sum(y_{\text{pred}}- y_{\text{obs}})^2}{\sum(y_{\text{obs}}-\overline{y}_\text{obs})^2},
\end{equation}
\noindent where the sum is over each timestep. PE values near 1.0 indicate perfect prediction, while negative values suggest the model performs worse than simply predicting the observed mean.

Frequency-domain validation employs Welch's power spectral density \cite{welch_use_1967} and magnitude-squared coherence (MSC) to quantify the linear relationship between measured and simulated signals:
\begin{equation}
\gamma_{xy}^2(f) = \frac{|S_{xy}(f)|^2}{S_{xx}(f)\,S_{yy}(f)},
\label{eq:S19}
\end{equation}
\noindent
where $S_{xy}(f)=E[X(f)Y^*(f)]$ is the cross-power spectral density and $S_{xx}(f)$ and $S_{yy}(f)$ are the auto-power spectral densities of the measured and simulated series. Welch's method segments the signals with overlap, applies a Hann window to each segment, computes the Fast Fourier Transform, and averages the resulting periodograms to estimate power spectral density.

\subsection{SECS magnetic-field interpolation: uncertainty and impact}\label{method:secs-validation}

Interpolation error in the magnetic driver propagates through the geoelectric calculation and into network currents. We assess this by comparing SECS-predicted horizontal fields with colocated magnetometers and reporting prediction efficiency and correlation.

\begin{figure}
\centering
\includegraphics[width=0.85\textwidth]{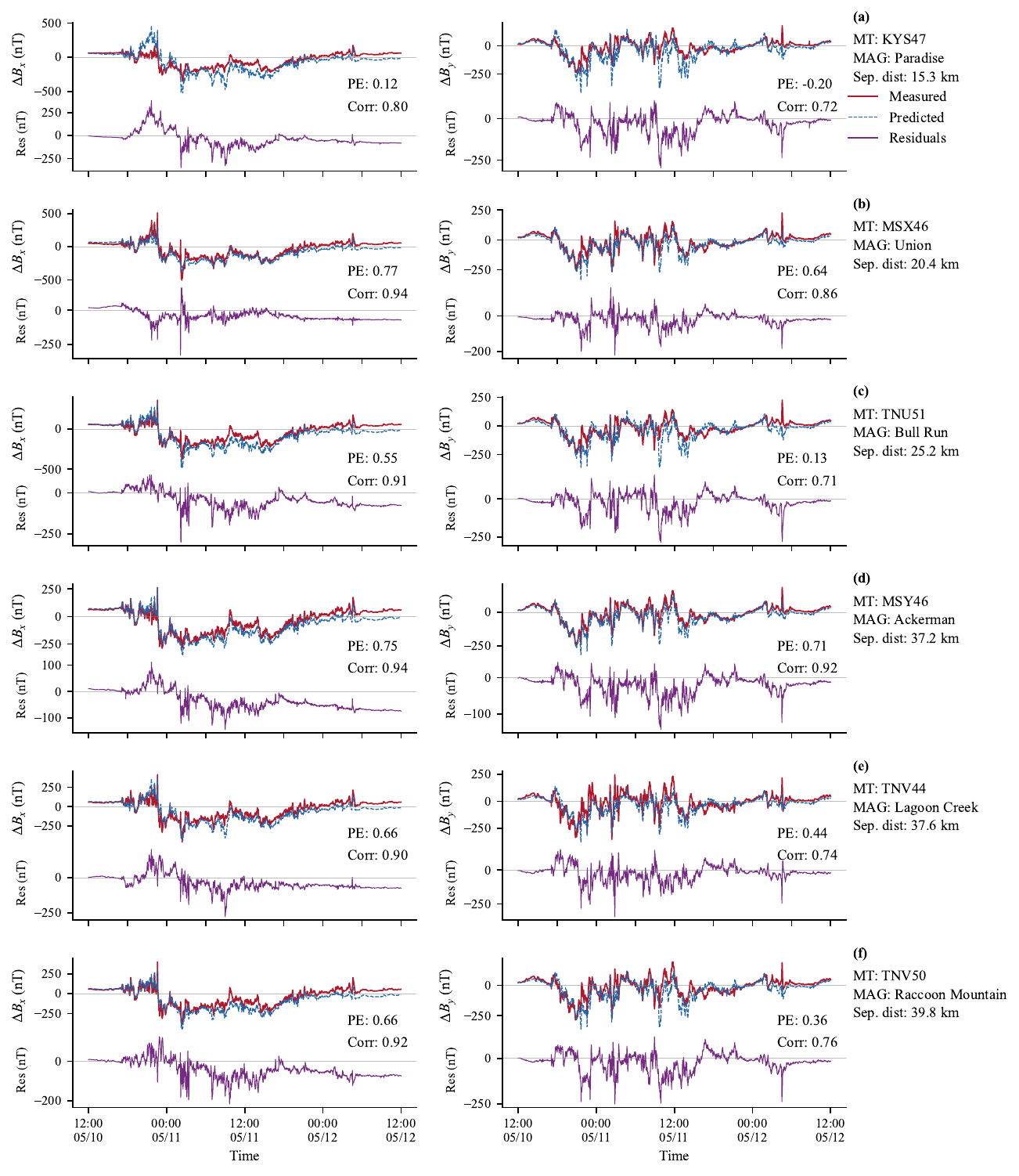}
\caption{\textbf{SECS geomagnetic field interpolation validation comparing predicted versus measured horizontal magnetic field components at magnetometer stations.} Residuals show the difference between predictions and measurements. Performance metrics include prediction efficiency (PE) and Pearson correlation coefficient (Corr). Sep. dist. denotes the separation distance in kilometers between the MT site used for prediction and the nearest magnetometer station.}
\label{fig:secs_mag_validation}
\end{figure}

Fig.~\ref{fig:secs_mag_validation} presents the performance of the SECS model in reproducing local geomagnetic field variations across the contiguous U.S. region. Each panel shows measured magnetometer data (solid lines) against SECS-predicted values (dashed lines) for both horizontal components ($\Delta B_x$ and $\Delta B_y$), with residuals plotted below. The prediction efficiency values range from $-0.20$ to $0.77$, with correlation coefficients between $0.71$ and $0.94$, indicating generally good agreement between the interpolated and measured magnetic field perturbations, despite site-to-site variations. The systematic negative residual drift observed during the recovery phase (May 11--12) is consistent with known SECS limitations at lower latitudes, where interpolation accuracy degrades due to sparse observatory coverage south of the Great Lakes \cite{rigler_interpolating_2019}.

\subsection{Measured--simulated GIC comparison (time and frequency domains)}\label{method:gic-timefreq}

\begin{figure}
\centering
\includegraphics[width=0.8\textwidth]{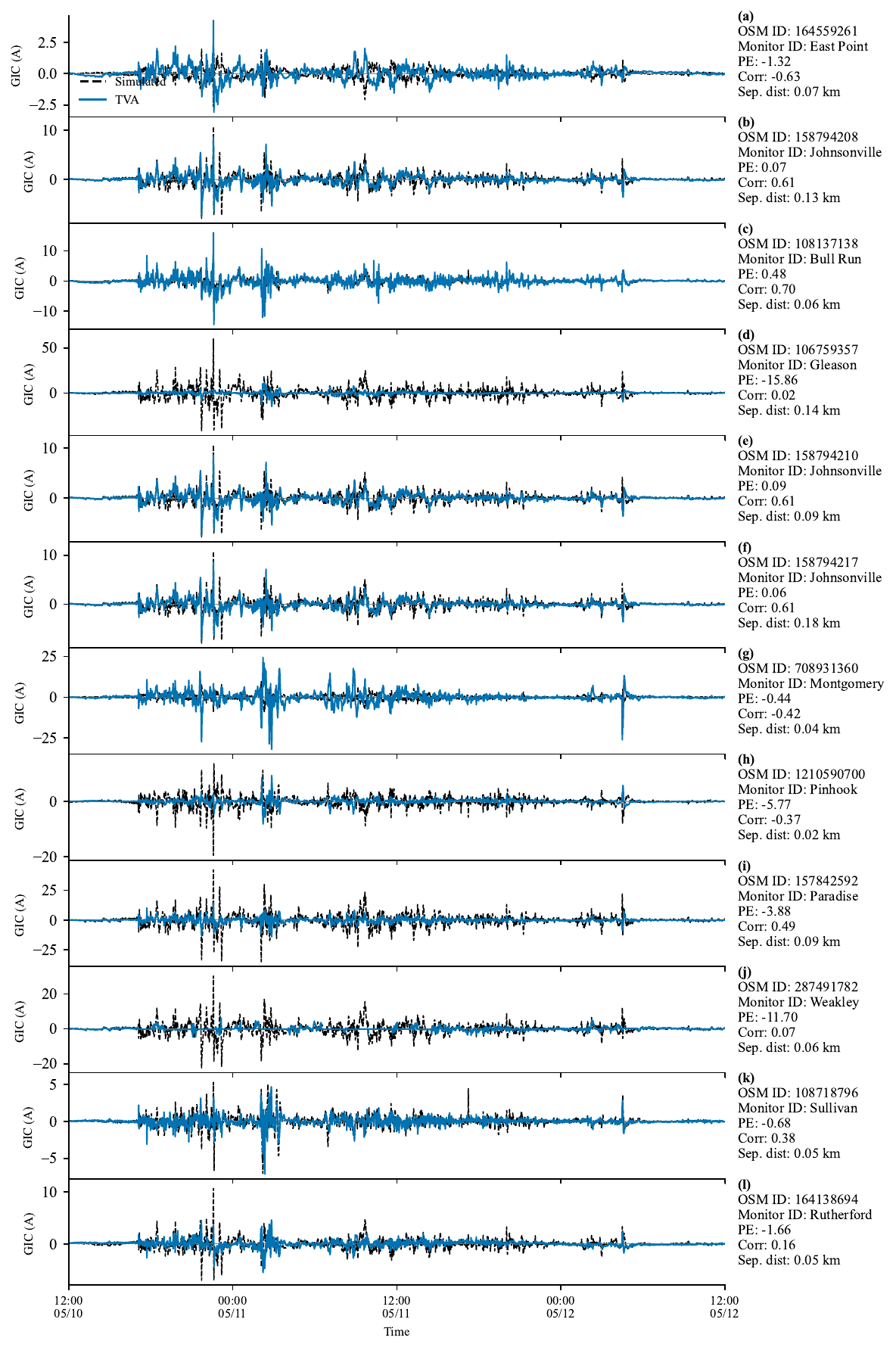}
\caption{\textbf{Time-domain comparison of measured (blue solid) versus simulated (black dashed) GIC values at TVA monitoring sites during the 10-12 May 2024 geomagnetic storm.}}
\label{fig:gic_time_comparison}
\end{figure}

Time-domain analysis of GIC signals presented in Fig.~\ref{fig:gic_time_comparison} reveals varying model performance across the TVA GIC monitoring network. The prediction efficiency ranges from highly negative values (PE $= -15.86$ at Gleason) to moderately positive values (PE $= 0.48$ at Bull Run). Correlation coefficients show generally good agreement when accounting for measurement polarity, with absolute correlation values ranging from $|r| = 0.02$ (Gleason) to $|r| = 0.70$ (Bull Run). Negative correlations, such as r = -0.63 at East Point, indicate strong anti-correlated behavior likely due to an inverted measurement device configuration rather than poor model performance. The model captures the general temporal evolution and peak magnitudes of GIC variations during the storm period, particularly on May 11th. Performance variations across sites reflect the complex interplay between regional electromagnetic forcing, local network topology, and substation-specific grounding configurations.

\begin{figure}
\centering
\includegraphics[width=0.7\textwidth]{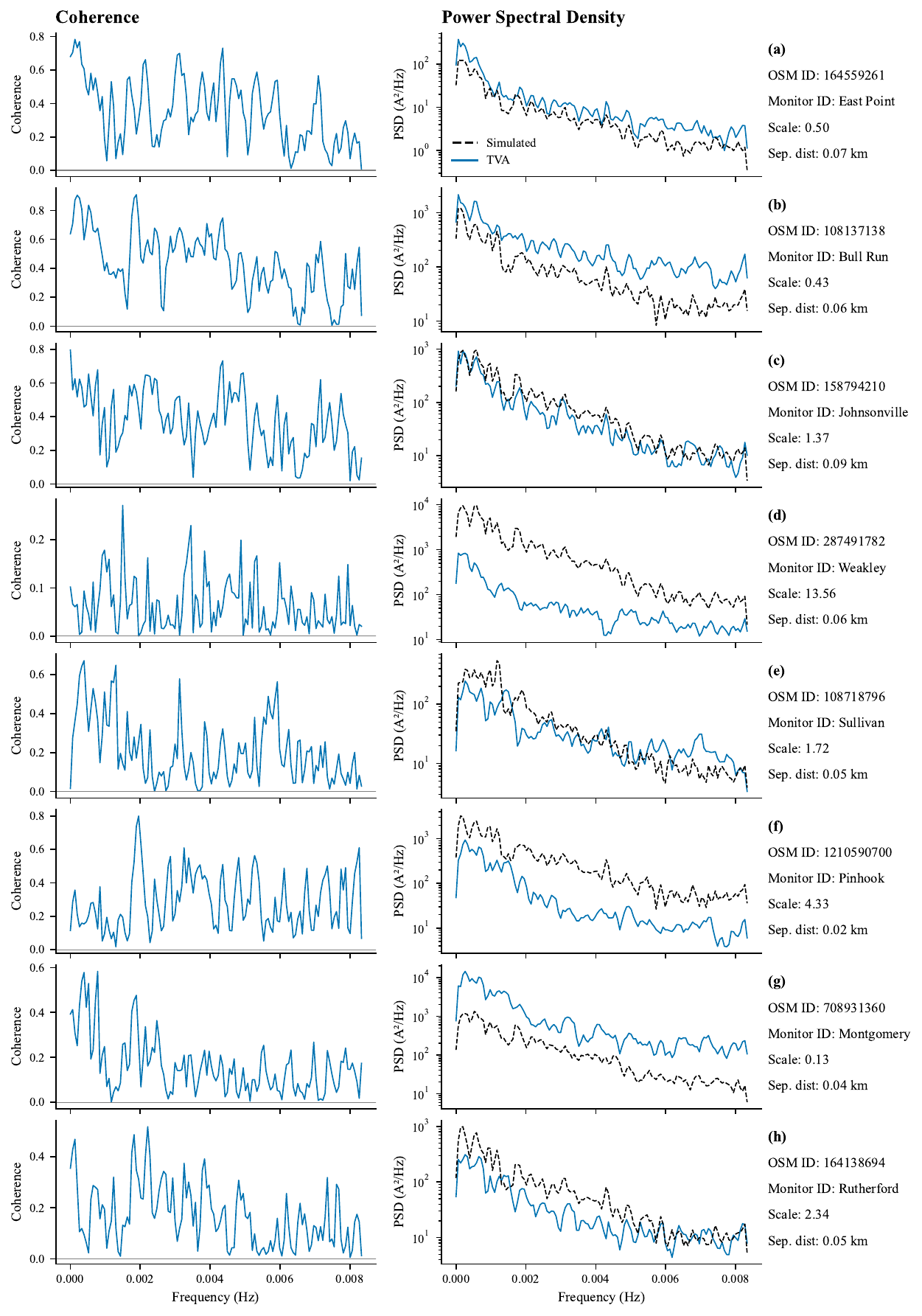}
\caption{\textbf{Frequency-domain analysis of GIC validation showing coherence (left panels) and Welch power spectral density estimates (right panels) for measured versus simulated GIC signals.} Coherence values close to 1 indicate strong frequency-by-frequency agreement between measured and simulated signals. Scale denotes the ratio of integrated simulated to measured power spectral density. Values greater than 1 indicate model overestimation of signal power, values less than 1 indicate underestimation. Sep. dist. denotes the separation distance in kilometers between the simulated substation and the nearest GIC monitoring device.}
\label{fig:gic_frequency_analysis}
\end{figure}

Fig.~\ref{fig:gic_frequency_analysis} presents the frequency-domain analysis using Welch power spectral density estimation and coherence analysis of the measured and simulated signals. The power spectral density plots reveal that both measured and simulated GIC signals exhibit similar spectral characteristics, with energy concentrated at lower frequencies ($< 0.004$ Hz) corresponding to periods longer than $\sim$ 4 minutes. The coherence analysis reveals a frequency-dependent correlation between the measured and simulated signals, with coherence values typically ranging from 0.2 to 0.8 across the frequency band, consistent with findings in \cite{weigel_evaluation_2019,oyedokun_frequency_2020}. The spectral agreement and coherence patterns demonstrate that the model effectively captures the dominant frequency content of GIC variations during geomagnetic disturbances.

\subsection{Alpha--beta regression versus physics-based simulation}\label{gic-pred-validation}

To validate GIC prediction capabilities, two fundamentally different approaches are compared: (1) an empirical alpha--beta scaling method following NERC TPL-007 protocols with enhanced spatial resolution, and (2) the physics-based simulation detailed in Section~\ref{method:lpm-app}. This comparison evaluates the relative strengths and limitations of data-driven versus first-principles approaches for assessing GIC risk.

The alpha--beta method represents the current industry standard for GIC assessment, relying on statistical relationships derived from limited historical observations. In contrast, the physics-based approach solves the complete electromagnetic and circuit physics without empirical approximations. Comparing these methods provides insights into the validity of simplified scaling approaches and identifies scenarios where more detailed modeling may be necessary.

The alpha--beta approach predicts maximum absolute GIC values using linear regression models trained exclusively on the May~2024 ``Gannon'' storm event \cite{wilkerson_gic--related_2025}. This data-driven methodology applies the NERC TPL-007 framework, incorporating higher spatial resolution ground conductivity models to enhance local accuracy.

The method employs two primary scaling factors. The geomagnetic latitude factor $\alpha(\lambda) = 0.001\,e^{0.115\,\lambda}$ accounts for the variation in geomagnetic field intensity with magnetic latitude $\lambda$, obtained through coordinate transformation from geographic to geomagnetic reference frames at the storm epoch. The ground conductivity factor $\boldsymbol{\beta}(x,y) = [\beta_x, \beta_y]$ captures local variations in subsurface electrical properties using magnetotelluric-derived 2-D conductivity models on an approximately $0.5^\circ$ spatial grid. These factors are spatially interpolated to substation locations using linear interpolation.

Multiple regression models were developed with varying input parameter combinations selected from the set $\{\alpha, \beta, \alpha\beta, \lambda\beta\}$. Each model was trained on measured GIC data from NERC ERO and TVA during the ``Gannon'' storm. Trained models were ranked based on their prediction performance. Only models with a correlation coefficient exceeding 0.75 were retained for ensemble predictions. Bootstrap sampling with 500 iterations combines individual model predictions to generate mean estimates with uncertainty bounds.

Return period extrapolation assumes a linear relationship between the geoelectric field magnitude and the resulting GIC. Scenarios for 1-in-100, 1-in-200, and 1-in-250~year events apply the scaling factor $s_{\text{rp}} = |\mathbf{E}|_{\text{rp}}/|\mathbf{E}|_{\text{Gannon}}$, where geoelectric field ratios are derived from the statistical analysis of historical geomagnetic events in Section~\ref{method:gmd-footprint-app}.

The validation employs regional aggregation rather than point-by-point comparison to address fundamental differences in spatial resolution and network representation between methods. The continental U.S. is divided into $200~\text{km} \times 200~\text{km}$ grid cells, with substations grouped by cell and mean absolute GIC values computed for each region. This approach minimizes the influence of local network topology variations while preserving regional spatial patterns in the distribution of GIC.

Both methods are validated against independently measured GIC data from TVA and NERC monitoring networks during the ``Gannon'' storm, providing ground truth for absolute prediction accuracy. The comparison encompasses four scenarios --- the reference ``Gannon'' event and three return period cases --- enabling assessment of methodological agreement across different geomagnetic disturbance intensities and the validity of linear scaling assumptions inherent in the alpha--beta approach.

\begin{figure}
    \centering
    \includegraphics[width=1\textwidth]{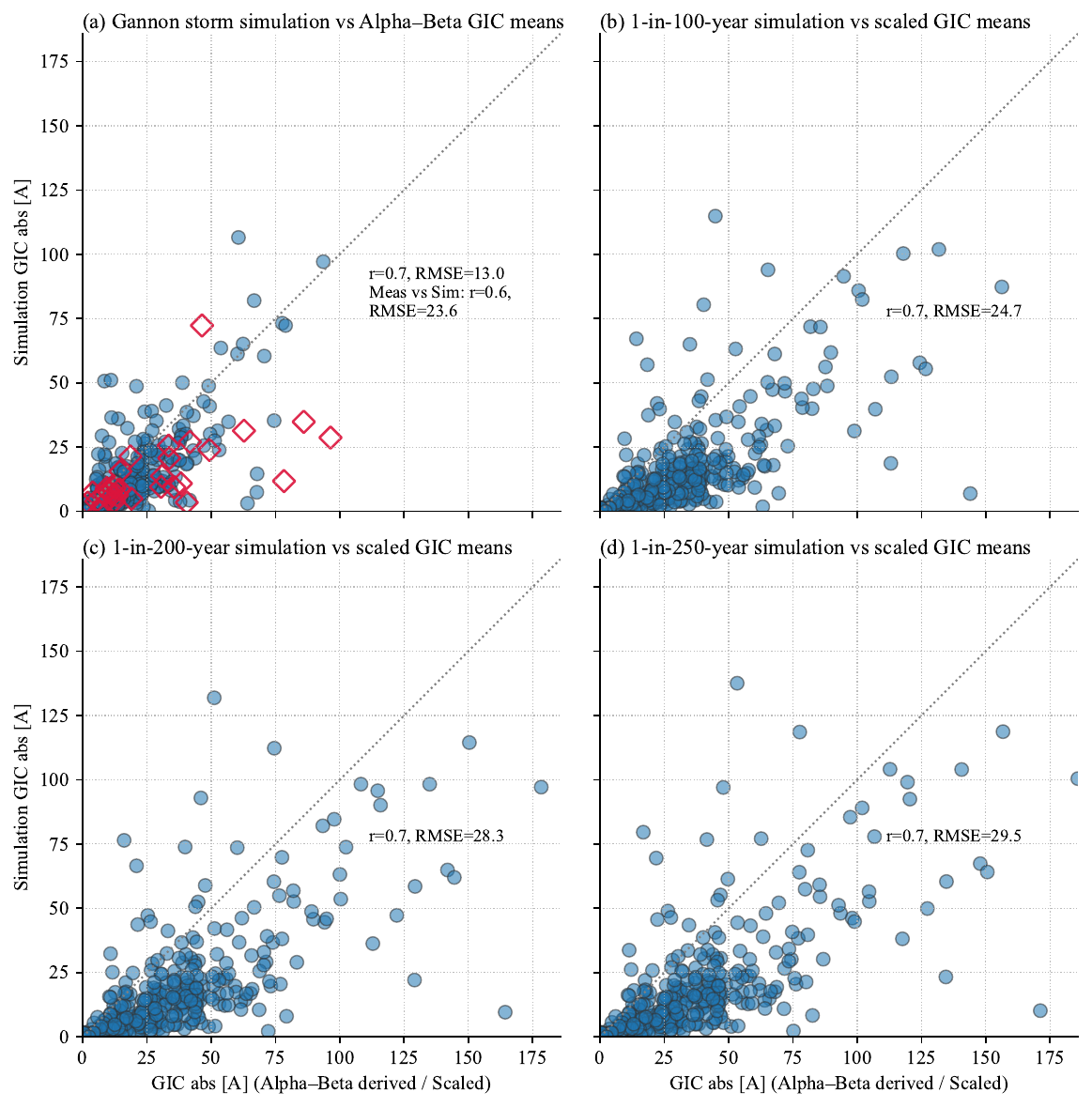}
    \caption{\textbf{Regional comparison of alpha--beta scaling versus physics-based simulation methods across storm intensities.} Crimson diamonds show measured GIC from TVA and NERC networks during the ``Gannon'' storm.}
    \label{fig:gic-validation-reg}
\end{figure}

Fig.~\ref{fig:gic-validation-reg} summarizes the regional comparison on 200~km cells. For the ``Gannon'' event, alpha--beta prediction regional means correlate well with the LPm simulation ($r=0.723$, RMSE$=15.5$~A, panel a), while independent measurements versus simulation show weaker agreement ($r=0.422$, RMSE$=29.2$~A). This weaker agreement reflects the inherent difficulty in predicting GIC, as it depends heavily on network configuration and grounding parameters. Across the return period scenarios, agreement between scaled alpha--beta and simulation remains stable (panels b--d: $r\approx0.72$--$0.75$, RMSE$\approx14$--$18$~A). The scatter structure reveals a dense, low-amplitude cluster and magnitude-dependent spread. At larger regional means, points fall below the 1:1 line, indicating that the scaled predictor tends to underestimate the upper tail relative to the physics model. We compare measurements only against the simulation in panel (a) to avoid training leakage, as the alpha--beta model was calibrated on the same ``Gannon'' event.

For applications that require rapid assessment and avoid the computational demands of running LPm simulations, the alpha-beta data-driven approach provides a practical alternative. For regional aggregation and sensitivity analysis --- such as identifying potential hotspots for mitigation planning --- alpha--beta scaling appears appropriate as it captures regional trends with reasonable fidelity. The method's computational efficiency makes it suitable for large-scale screening studies where precise point estimates are less critical than understanding relative exposure patterns. While it remains difficult to definitively assess whether the physics model systematically over- or underpredicts without extensive validation data, both approaches capture the observed trends, with the physics-based method providing more detailed spatial resolution at the cost of increased computational complexity.

\section{Supply-side view of economic loss with the Ghosh model}\label{sec:ghosh-econ-model}

\begin{figure}
	\centering
	\includegraphics[width=1\textwidth]{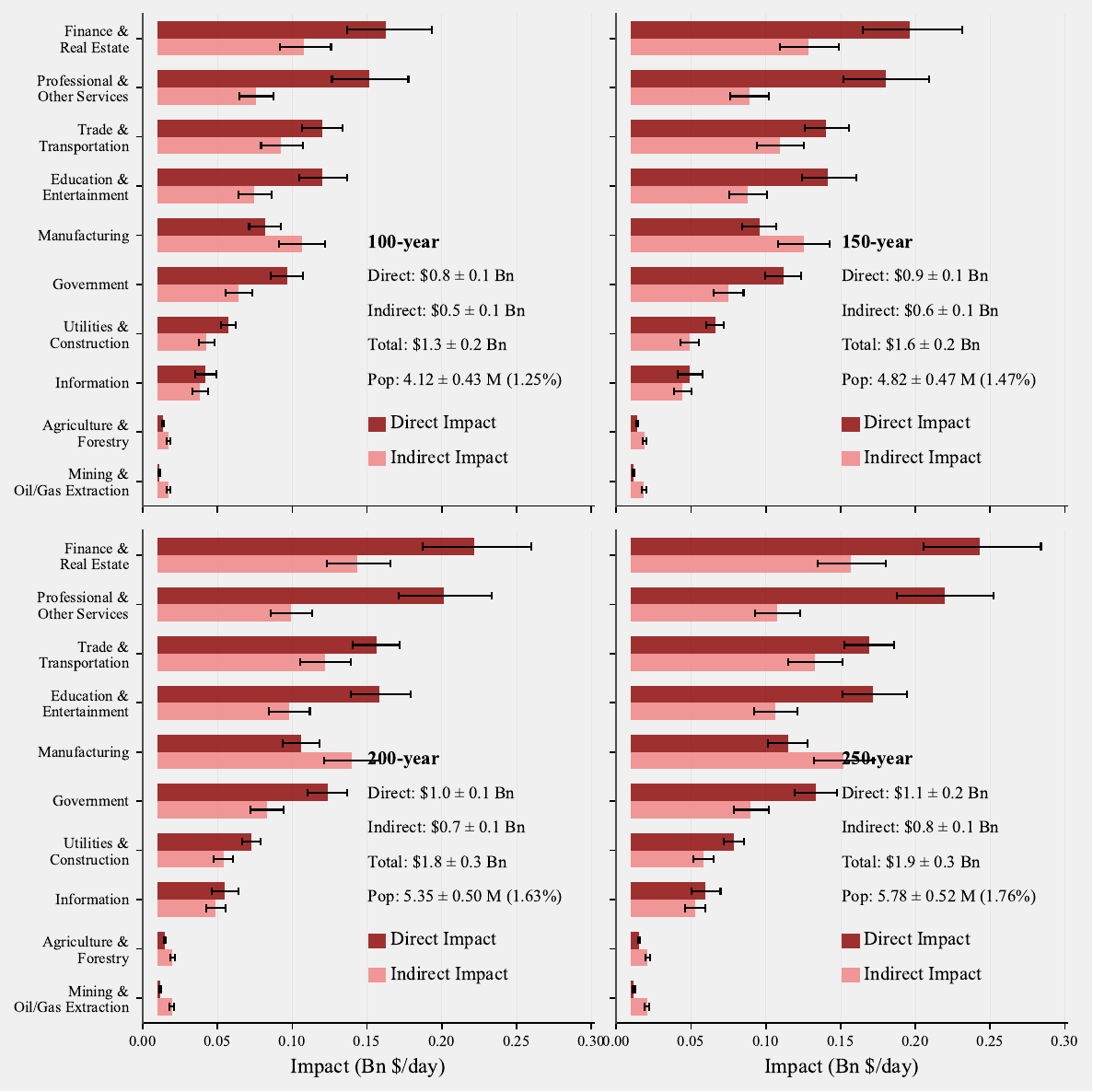}
	\caption{\textbf{Sectoral breakdown of direct and indirect economic impacts under the Ghosh model.} Losses concentrate in sectors with strong forward linkages, in contrast to the Leontief results in Figure 6 of the main text.}
	\label{fig:io_comparison_ghosh}
\end{figure}

The Ghosh and Leontief models are compared on the same set of failure scenarios to assess the sensitivity of total impacts to the choice of propagation framework. Total daily losses agree closely across all return periods, indicating that the total impacts are not sensitive to the choice of supply-driven versus demand-driven propagation. Figure~\ref{fig:io_comparison_ghosh} shows the Ghosh results broken down by sector, where indirect losses exceed direct losses in some cases.

The two models disagree on the sectoral breakdown. Under the Ghosh model, losses are concentrated in sectors with strong forward linkages, including finance and real estate, and professional and other services. The sectoral attribution of the two models for a 250-year storm is compared in Table~\ref{tab:io_comparison}. For the Leontief model, losses concentrate in upstream supply-chain sectors such as manufacturing, education, and entertainment, because consumer demand cascades upstream through long supply chains. This divergence reflects the direction of propagation through the input--output system.





\begin{table}
\centering
\caption{\textbf{Power system network and LPm method parameters for Monte Carlo simulation.}
The simulation accounts for uncertainties in transformer configuration, winding resistances, line resistances, grounding, and GIC blocking devices.}
\label{tab:power_system_params}
\vspace{2mm}
\footnotesize
\setlength{\tabcolsep}{3pt}
\renewcommand{\arraystretch}{0.95}

\begin{tabular*}{0.9\linewidth}{@{}l @{\extracolsep{\fill}} l@{}}
\hline
\textbf{Parameter} & \textbf{Value/Distribution} \\
\hline
\multicolumn{2}{@{}l}{\textit{Transformer Configuration}} \\
Number per substation & 1--3 (type-dependent)$^*$ \\
Type assignment & Based on voltage ratio \\
\hline
\multicolumn{2}{@{}l}{\textit{Transformer Winding Resistances ($\Omega$/phase)}} \\
GY-GY-D & Pri: 0.2,\; Sec: 0.1 \\
GY-GY & Pri: 0.04,\; Sec: 0.06 \\
Auto & Series: 0.04,\; Common: 0.06 \\
GSU & Pri: 0.15,\; Sec: $\infty$ \\
GSU w/ GIC BD & Pri: 0.1,\; Sec: $\infty$ \\
GY-D & Pri: 0.05,\; Sec: $\infty$ \\
\hline
\multicolumn{2}{@{}l}{\textit{Transmission Line Resistances ($\Omega$/km/phase)}} \\
765 kV & 0.01 \\
500 kV & 0.0141 \\
345 kV & 0.0283 \\
230 kV & 0.05 \\
161 kV & 0.08 \\
Length adjustment & $\times$1.03 (sag/meander) \\
\hline
\multicolumn{2}{@{}l}{\textit{Substation Grounding}} \\
Grounding resistance & Uniform$^\dagger$ [0.1, 0.2] $\Omega$ \\
Ungrounded probability & 1\% \\
\hline
\multicolumn{2}{@{}l}{\textit{GIC Blocking Devices}} \\
Line blocking probability & 1\% \\
Effect on blocked lines & $R \rightarrow \infty$, sources = 0 \\
\hline
\multicolumn{2}{@{}l}{\textit{Monte Carlo Settings}} \\
Number of simulations & 2000 \\
\hline
\end{tabular*}

\vspace{2mm}
\medskip
\centering
\begin{minipage}{0.9\linewidth}
\footnotesize
$^*$Type-dependent: 1---2 for GSU, 1---3 for transmission substations.\\
$^\dagger$Uniform: Continuous uniform distribution sampled between specified values.\\
\textit{Abbreviation:} Pri = primary; Sec = secondary.
\end{minipage}
\end{table}

\begin{table}
\centering
\caption{\textbf{Transformer failure fragility and reliability analysis parameters.}
Failure probability is modeled using lognormal fragility curves with age-dependent capacity degradation. Monte Carlo simulation continues until 5\% convergence tolerance is achieved for population impact estimates.}
\label{tab:reliability_params}
\vspace{2mm}
\footnotesize
\setlength{\tabcolsep}{3pt}
\renewcommand{\arraystretch}{0.95}

\begin{tabular*}{0.9\linewidth}{@{}l @{\extracolsep{\fill}} l@{}}
\hline
\textbf{Parameter} & \textbf{Value/Distribution} \\
\hline
\multicolumn{2}{@{}l}{\textbf{Fragility Curve Parameters}} \\
Median capacity $\theta_0$ & 75 A/phase \\
Lognormal dispersion $\beta$ & Uniform$^\dagger$ [0.25, 0.50] \\
GIC prediction uncertainty & Uniform$^\dagger$ [0.6, 1.4] \\
\hline
\multicolumn{2}{@{}l}{\textbf{Age-Related Degradation}} \\
Transformer age distribution & 55\%: Uniform$^\dagger$ [33, 50] years \\
 & 45\%: Uniform$^\dagger$ [1, 32] years \\
Weibull shape parameter $\beta_{\text{age}}$ & Uniform$^\dagger$ [1, 3] \\
Weibull scale parameter $\eta_{\text{age}}$ & Uniform$^\dagger$ [30, 50] years \\
Capacity reduction factor & $\theta = \theta_0 (1 - 0.6 \times F_{\text{age}})$ \\
\hline
\multicolumn{2}{@{}l}{\textbf{Convergence Criteria}} \\
Tolerance & 5\% relative half-width (95\% CI) \\
Minimum iterations & 5000 \\
Maximum iterations & User-defined \\
\hline
\end{tabular*}

\vspace{2mm}
\medskip
\centering
\begin{minipage}{0.9\linewidth}
\footnotesize
$^\dagger$Uniform: Continuous uniform distribution sampled between specified values.
\end{minipage}
\end{table}

\begin{table}
\centering
\caption{\textbf{Sectoral breakdown of total daily losses under the Leontief and Ghosh models for a 250-year storm.} Values are in \$M/day at mean confidence. The two models agree on headline total to within six percent but disagree on sectoral attribution, reflecting the direction of propagation through the input--output system.}
\label{tab:io_comparison}
\vspace{2mm}
\footnotesize
\setlength{\tabcolsep}{3pt}
\renewcommand{\arraystretch}{0.95}
\begin{tabular*}{0.9\linewidth}{@{}l @{\extracolsep{\fill}} rrr@{}}
\hline
\textbf{Sector} & \textbf{Leontief} & \textbf{Ghosh} & \textbf{Difference} \\
\hline
Manufacturing                   & $-607.8$ & $-246.9$ & $-360.9$ \\
Finance and real estate         & $-381.7$ & $-380.0$ & $-1.7$ \\
Education and entertainment     & $-299.8$ & $-258.5$ & $-41.3$ \\
Professional and other services & $-251.3$ & $-307.5$ & $+56.2$ \\
Government                      & $-186.9$ & $-203.7$ & $+16.8$ \\
Information                     & $-75.2$  & $-92.7$  & $+17.5$ \\
Utilities and construction      & $-75.0$  & $-116.9$ & $+41.9$ \\
Agriculture                     & $-72.1$  & $-16.6$  & $-55.5$ \\
Mining                          & $-49.4$  & $-12.6$  & $-36.8$ \\
Trade and transportation        & $-39.2$  & $-282.1$ & $+242.9$ \\
\hline
\textbf{Total}                  & $\mathbf{-2038.3}$ & $\mathbf{-1917.6}$ & $\mathbf{-120.7}$ \\
\hline
\end{tabular*}
\end{table}

\clearpage
\bibliography{ref}


%
%
%
%
%

\end{document}